\documentclass{nature}
\usepackage[left=2.5cm, right=2.5cm,top=2.5cm,bottom=3cm]{geometry}
\newgeometry{top=1cm, bottom=2cm, left=2cm, right=2cm}

\usepackage{graphicx}
\usepackage[font=normal, skip=0pt, labelfont=bf]{caption}

\usepackage{bm}

\usepackage{xspace}
\usepackage{booktabs}

\usepackage{amsmath, amssymb}
\usepackage{times}
\usepackage{multirow}

\usepackage[parfill]{parskip}
\usepackage{xcolor}
\definecolor{mplred}{HTML}{D62728}

\usepackage{hyperref}
\hypersetup{breaklinks, plainpages=false, colorlinks=true, anchorcolor=blue!50!black, citecolor=blue!50!black, linkcolor=blue!50!black, urlcolor=mplred, bookmarks=false}

\makeatletter

\let\saved@includegraphics\includegraphics
\AtBeginDocument{\let\includegraphics\saved@includegraphics}
\makeatother

\renewcommand{\figurename}{\textbf{Figure}}

\usepackage[caption=false]{subfig}
\makeatletter
\renewcommand\setcaptiontype[1]{\edef\@captype{#1}\ignorespaces}
\makeatother

\usepackage{longtable}
\usepackage{supertabular}
\usepackage[flushleft]{threeparttable}

\makeatletter
\def\@mb@citenamelist{cite,citep,citet,citen,citealp,citealt,citepalias,citetalias}
\makeatother
\usepackage{multibib}
\newcites{methods}{\Large\bf References (Methods):\bigskip}

\makeatother

\title{Selection bias obfuscates the discovery of fast radio burst sources}

\newcommand{\MB}{\href{https:/orcid.org/0000-0002-3615-3514}{\textcolor{blue!50!black}{Mohit~Bhardwaj}}}
\newcommand{\JL}{\href{https:/orcid.org/0009-0003-8484-2925}{\textcolor{blue!50!black}{Jimin~Lee}}}
\newcommand{\KJ}{{Kevin~Ji}}

\newcommand{\CMU}{McWilliams Center for Cosmology, Department of Physics, Carnegie Mellon University, Pittsburgh, PA 15213, USA}

\author{\MB\textsuperscript{\textdagger}$^{1}$, \JL\textsuperscript{\textdagger}$^{1}$, \KJ$^{1}$}

\begin{document}

\maketitle

\begin{affiliations}
\item \, \CMU \\
\textsuperscript{\textdagger}These authors contributed equally to this work
\end{affiliations}


\begin{abstract}

\textbf{
Fast Radio Bursts (FRBs) are a newly discovered class of extragalactic radio transients characterised by their high energy and short duration ($\sim \mu$s$-$ms)\cite{Lorimer+2007}. Their elusive physical origin remains a subject of ongoing research, with magnetars emerging as leading candidates\cite{CHIME+2020a,Bochenek+2020c}. Previous studies have employed various methodologies to address the FRB origin problem, including demographic analyses of FRB host galaxies and their local environments\cite{Li2020ApJ,Gordon2023ApJ,2023arXiv231010018B}, assessments of FRB rate evolution with redshift\cite{Luo2020MNRAS,Hashimoto2020MNRAS,Zhang2021MNRAS}, and searches for proposed multi-messenger FRB counterparts\cite{2021Univ....7...76N}. However, these
studies are susceptible to significant biases stemming from unaccounted radio and optical selection effects.
Here we present empirical evidence for a substantial selection bias against detecting FRBs in galaxies with large inclination angles (edge-on) using a sample of hosts identified for FRBs discovered by untargeted surveys. This inclination-related bias likely leads to a significant underestimation (by about a factor of two) of the FRB rates reported in the literature and disfavours globular clusters as the dominant origin of FRB sources, as previously speculated\cite{2023arXiv231010018B}. These conclusions have important implications for FRB progenitor models and targeted FRB follow-up strategies. We further investigate the impact of this bias on the relative rate of FRBs in different host environments. Our analysis suggests that scattering in FRB hosts is likely responsible for the observed bias\cite{Chawla2022ApJ,Ocker2022ApJ}. However, a larger sample of localised FRBs is required to robustly quantify its contribution in the inclination-related selection bias.}

\end{abstract}

\bigskip

In this study, we investigate potential biases introduced by the inclination angle ($i$) of FRB hosts on the detectability of FRBs in various radio surveys. We select published FRBs localised with arcsecond precision and associated with their hosts with a PATH (Probabilistic Association of Transients to their Hosts)\cite{Aggrawal2021ApJ} posterior probability $>$ 95\%. To minimise biases from targeted surveys, we focus on hosts identified from two untargeted surveys: the Commensal Real-time ASKAP Fast Transients Survey (CRAFT)\cite{craft2010PASA} and Deep Synoptic Array (DSA)-110\cite{Law2023}.

Our study is limited to non-repeating FRBs, as most repeating FRBs have been identified by the Canadian Hydrogen Intensity Mapping Experiment (CHIME)\cite{2023ApJ...947...83C}, and their precise localization ($\lesssim$ few arcseconds) through targeted follow-up campaigns excludes them from our analysis. We also restrict our sample to hosts with an apparent r-band magnitude (m$_{r}$) $<$ 21 AB to ensure robust inclination angle measurements using archival imaging survey data.

The imaging surveys used in this analysis are the Panoramic Survey Telescope and Rapid Response System (Pan-STARRS) survey\cite{2016arXiv161205560C} and the DESI (Dark Energy Spectroscopic Instrument) survey\cite{2016arXiv161100036D}, which are sensitive to galaxies with m$_{r}$ $\lesssim$ 22 AB mag and 23.5 AB mag, respectively. We prioritise DESI survey data due to its greater depth, using Pan-STARRS data only when the FRB host is outside the DESI footprint. This approach yields a sample of 23 FRB hosts, listed in Table \ref{tab:host_universe_frbs}.

To estimate the inclination angles ($i$) of the 23 FRB hosts, we employ two methods: parametric profile fitting using the software \texttt{AutoProf}\cite{Stone2021b} and a non-parametric elliptical isophotal fitting routine from the software \texttt{Photutils}\cite{larry_bradley_2023_7946442}. The $i$ estimates from both methods are reported in Table \ref{tab:host_universe_frbs}, and their consistency supports the robustness of our calculations (Methods). Notably, the best-fitted Sérsic index from the first method is $\leq$ 1 for all 23 FRB host galaxies, indicating that these hosts are likely disk-dominated late-type galaxies\cite{yuan+2021}, as also found in a recent study focusing on local Universe FRB hosts ($z<0.1$)\cite{2023arXiv231010018B}. However, it is important to note that at least two FRBs, both repeating sources, have been localised to irregular star-forming dwarf galaxies\cite{Chatterjee+2017,2022Natur.606..873N}. These FRB sources are situated within the interstellar medium (ISM) of their respective hosts, coinciding with regions of active star formation.

We next compare the observed distribution of $i$ for FRB hosts with that expected from a sample of randomly selected field galaxies. Our null hypothesis asserts that the inclination distribution of FRB hosts mirrors that of field galaxies. Given that galaxies are viewed at random angles and that there is no preferred orientation in the Universe\cite{Saadeh2016PhRv}, we expect a uniform distribution of the cosine of inclination angles ($\cos(i)$) among field galaxies\cite{Joachimi2015SSRv}. However, deviations from this expected uniformity are observed, which can be attributed to the limitations inherent in the methods used to estimate $i$ (Methods).

To address these limitations and ensure a fair comparison with our FRB host sample, which are disk-dominated, we derived the expected $i$ distribution of 23 randomly sampled disk-dominated late-type galaxies from the Sloan Digital Sky Survey (SDSS) DR16 catalogue\cite{SDSSDR16_2020}. To exclude early-type galaxies, we apply the following u$-$r colour criterion: $u - r < 2.3$\cite{Strateva2001AJ}. Additionally, we use the same constraint on apparent r-band magnitude (m$_{\rm r}$ $<$ 21 AB mag) as employed in selecting FRB hosts. We note that these selection criteria do not affect the conclusions of our study (Methods). For the selected galaxies, we utilise morphological parameters provided by the SDSS DR16 catalogue, including the axial-ratio ($b/a$), calculated by fitting an exponential disk profile to the galaxy light, to determine their $i$.

\begin{figure*}
\centering
    \includegraphics[scale=0.60]{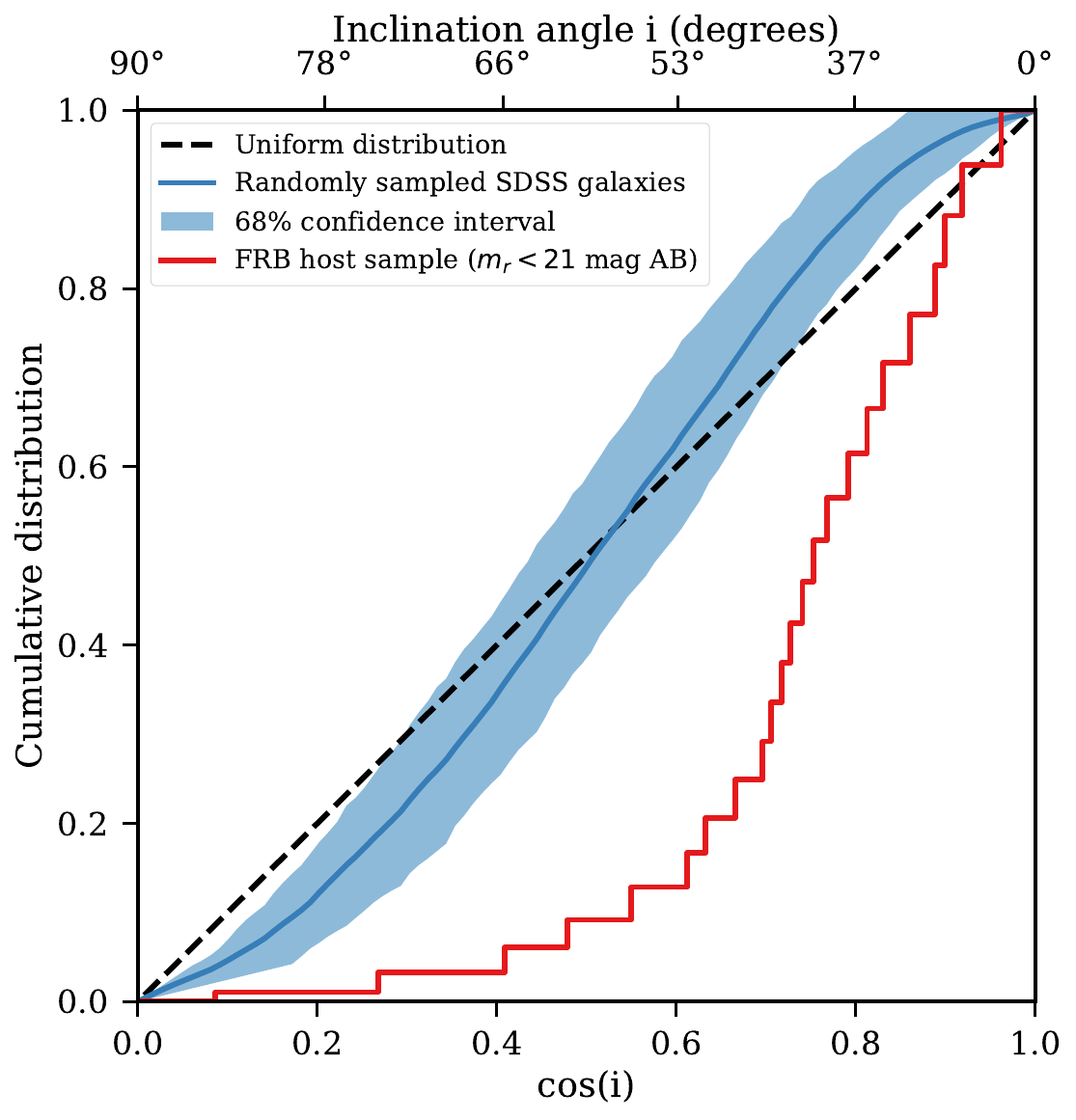}
    \caption{The cumulative distribution function (CDF) of the cosine of the inclination angles measured for 23 FRB hosts in our sample is compared with that of randomly sampled disk-dominated late-type galaxies from SDSS. The blue shaded region represents the 68\% credible bound on the SDSS galaxy CDF to account for the small sample size.}
    \label{fig:main_cdf}
\end{figure*}

The cumulative distribution functions (CDFs) of the $\cos(i)$ values for both the randomly sampled SDSS galaxies and FRB hosts are displayed in Figure \ref{fig:main_cdf}. To ensure a fair comparison, we employ $i$ estimates derived from the parametric profile fitting method, which aligns with the technique used in the SDSS pipeline. Our findings remain consistent even if $i$ estimates from the non-parametric elliptical isophotal fitting method are used. Additionally, to mitigate potential sampling effects, we generate 10,000 realisations of a sample comprising 23 randomly selected SDSS disk-dominated galaxies with m$_{r} <$ 21 AB mag. From these realisations, we estimate a 68\% credible bound on the SDSS galaxy CDF, also illustrated in Figure \ref{fig:main_cdf}.

To compare the $i$ distributions of the FRB host sample with randomly sampled SDSS galaxies, we use two non-parametric tests: the two-sample Anderson-Darling (AD) test\cite{scholz1987k} and the Mann-Whitney (MW) U test\cite{mann1947test}. The p-values obtained, $0.0003$ from the AD test and $0.00012$ from the MW test, signify a significant difference in the $\cos(i)$ distributions between the localised FRBs and the SDSS galaxy sample. Hence, we conclude that the $i$ distribution of our FRB host sample distinctly differs from that of field galaxies in the Universe.


An alternative approach to assess the difference in CDFs involves dividing the observed mean CDF of SDSS galaxies into bins. We opt for four bins, which ensures at least one FRB host galaxy per bin. Note that increasing the number of bins would result in empty bins, which would only strengthen our conclusion. These bins correspond to the quartiles of the $\cos{i}$ histogram for the SDSS galaxies, ensuring a uniform distribution of $\cos{i}$ among these randomly sampled galaxies.

\begin{figure*}
\centering
    \subfloat{\includegraphics[width=0.5\textwidth]{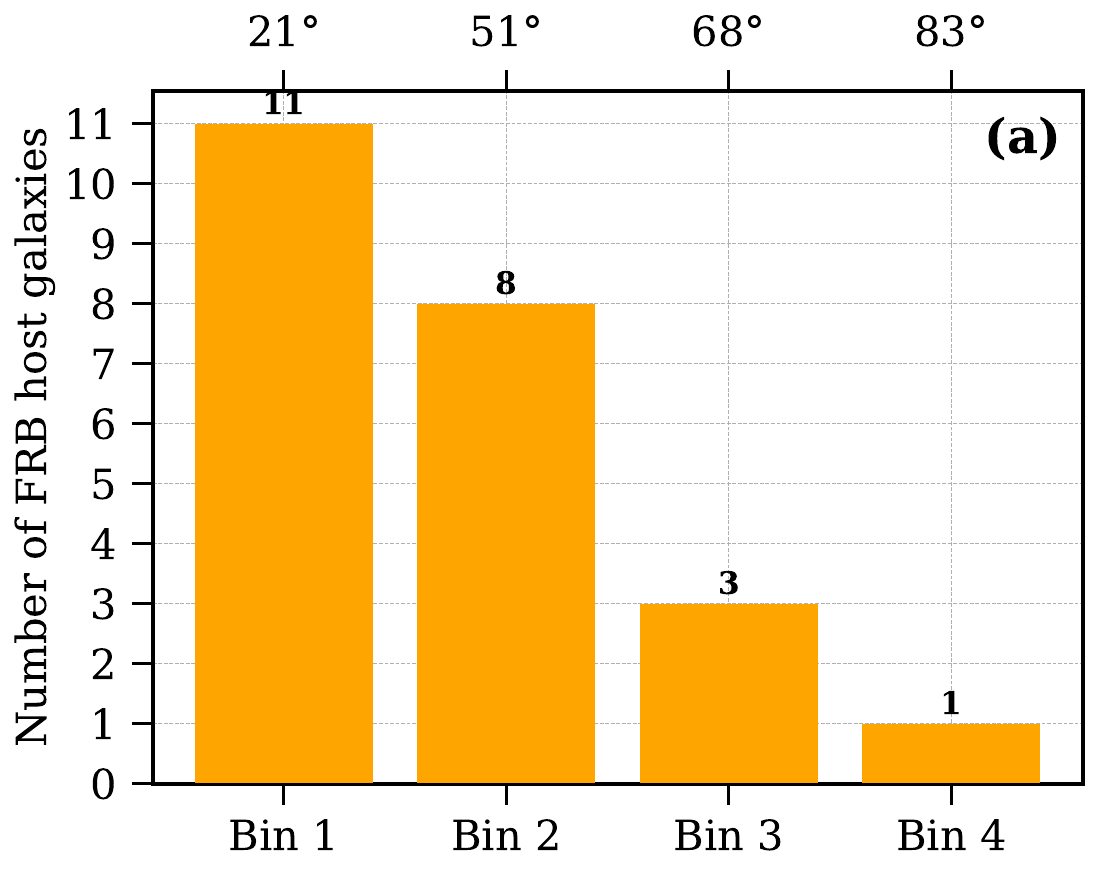}}\\
    \subfloat{\includegraphics[width=0.42\textwidth]{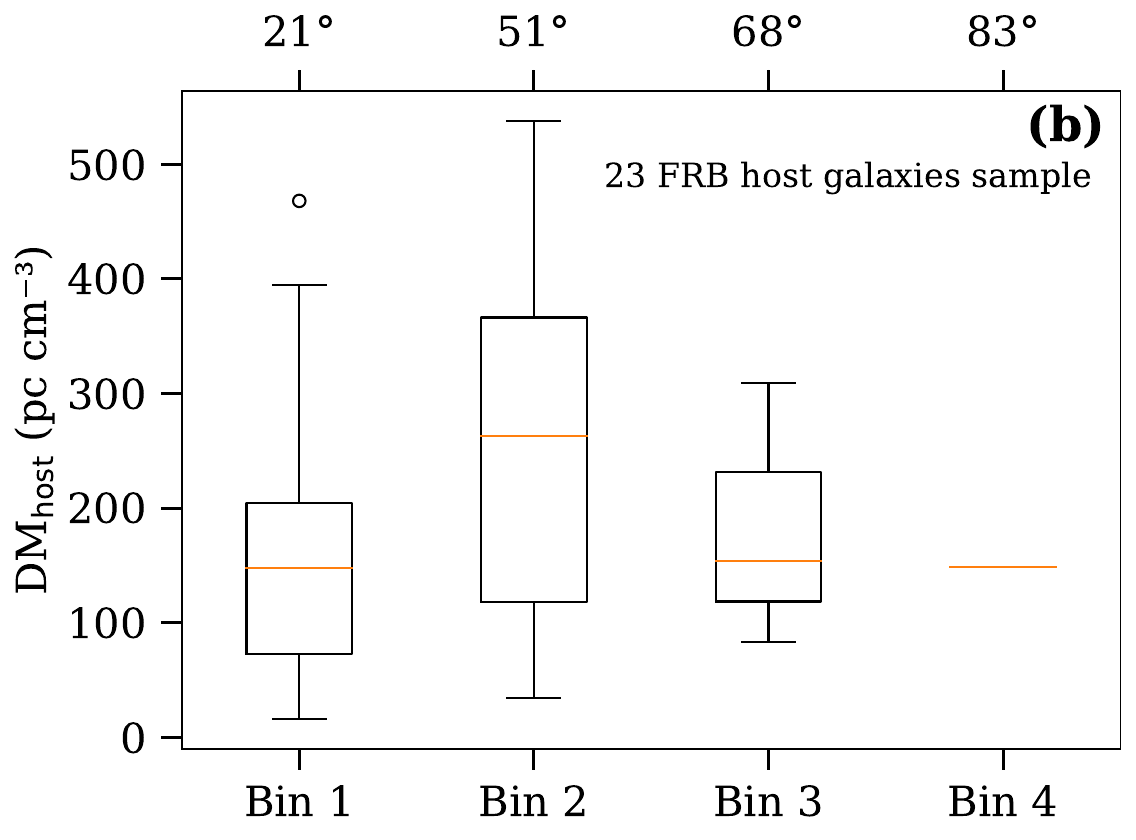}}
    \subfloat{\includegraphics[width=0.42\textwidth]{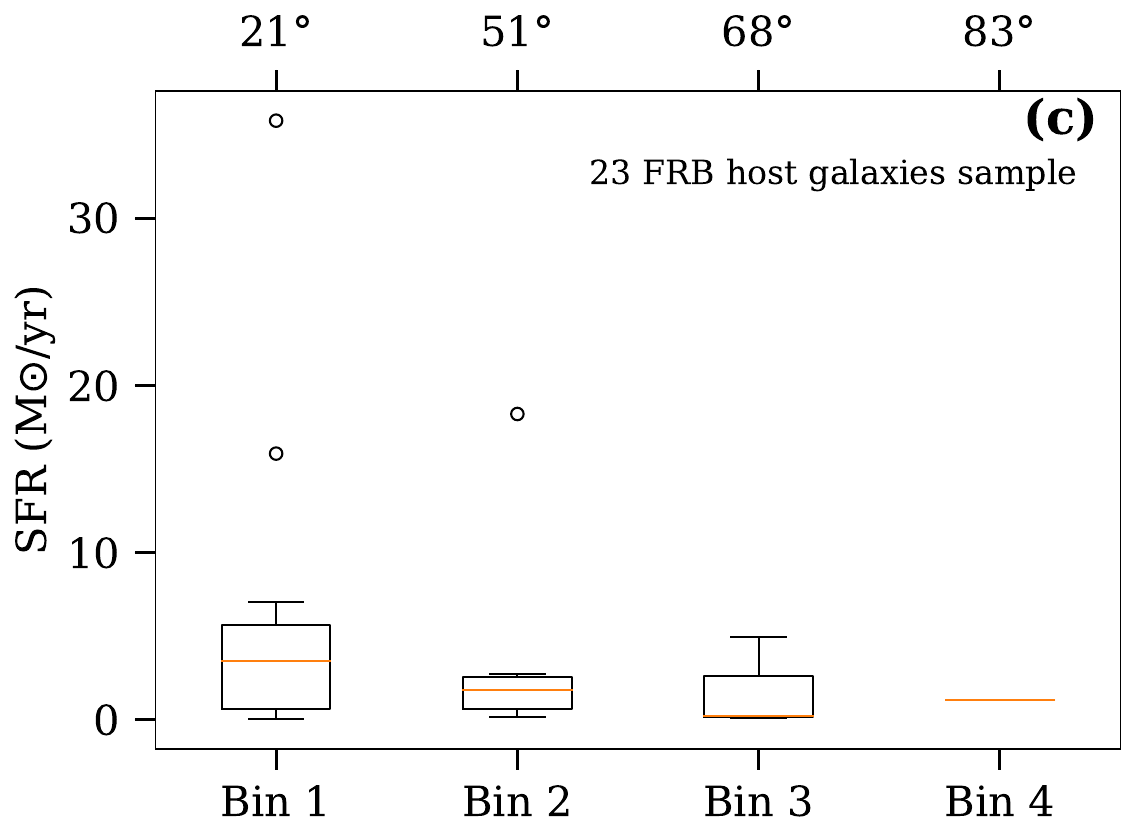}}\\
    \subfloat{\includegraphics[width=0.42\textwidth]{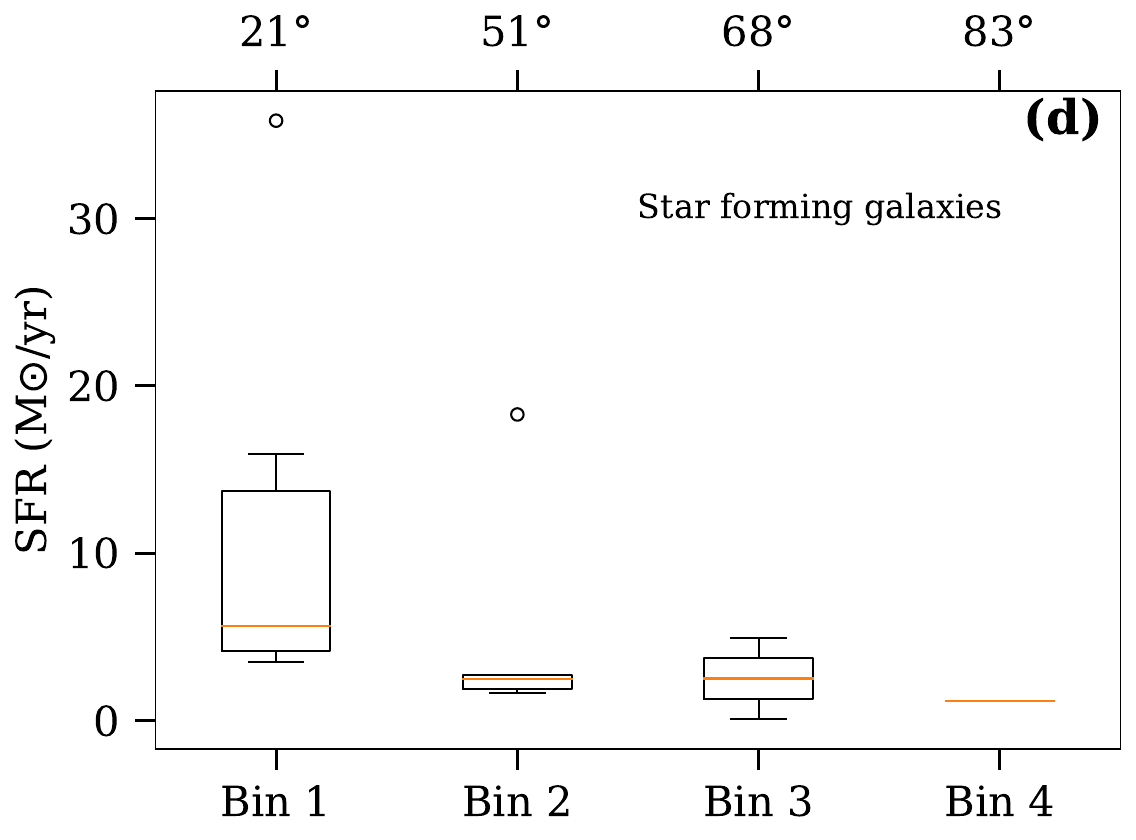}}
    \subfloat{\includegraphics[width=0.42\textwidth]{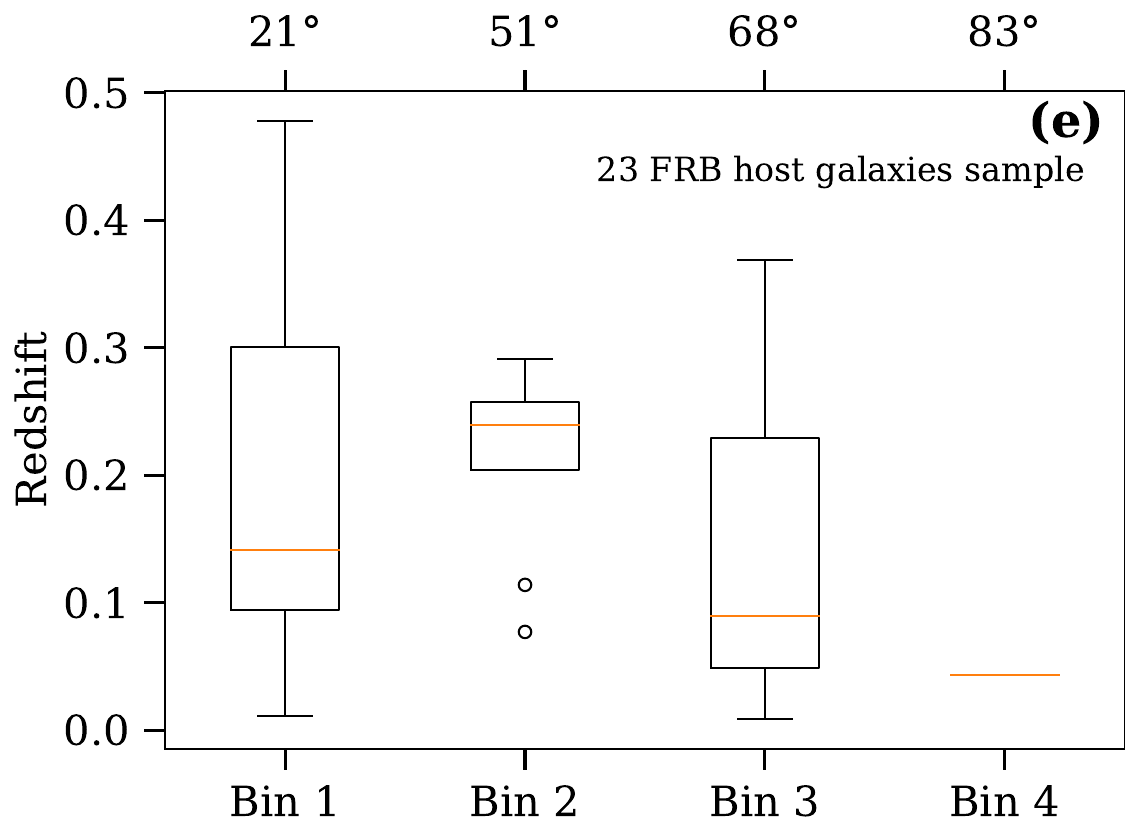}}\\
    \caption{
    (a) Histogram displays the number of FRB host galaxies within each $i$ bin as detailed in Table \ref{tab:bin_range}. (b) Box plot shows the DM$_{\rm host}$ range of FRB hosts within each $i$ bin. Similarly, (c) illustrates the SFR range of FRB hosts. (d) Presents the SFR range of FRB hosts classified as star-forming using the criterion: log($\rm sSFR$) $> -10$\cite{2004MNRAS.351.1151B}. Finally, (e) showcases the redshift range of FRB host galaxies. All relevant values are listed in Table \ref{tab:host_universe_frbs}.
    In each box plot, the orange lines depict the median values. The box edges indicate the lower and upper quartiles, covering 25\% of the data at each end. The ends of the horizontal lines show the maximum and minimum data values, and outliers are represented as individual points on the diagram.}  
    \label{fig:boxplots}
\end{figure*}

The $i$ range for each bin is listed in Table \ref{tab:bin_range}. In Figure \ref{fig:boxplots}a, we illustrate the distribution of FRB hosts across the four bins, which does not follow a uniform distribution. The majority of FRBs (48\% of the FRB hosts) have $i< 42^{\circ}$, and there is an evident scarcity of edge-on galaxies (only one with $i > 75^{\circ}$). This inclination-related bias is purely observational, as there is no inherent reason for FRB sources to be preferentially located in galaxies appearing nearly face-on to us, suggesting a significant underestimation of FRB all-sky rates. Based on the number of FRB hosts in Bin 1 (11) and disregarding differences in exposure and sensitivity between DSA-110 and CRAFT surveys, this underestimation (44 expected instead of 23 observed) can be a factor of two assuming Poisson statistics. A more robust estimate would require a larger sample of FRB hosts and must consider the various systematic factors present in each survey. 

This has two major implications: First, it reinforces previous assertions made using the reported all-sky rate in the literature, suggesting that the majority of FRBs are likely repeating sources\cite{2019NatAs...3..928R,Bhardwaj+2021b,2023PASA...40...57J}. Second, any targeted follow-up campaigns aiming to determine the relative rates of various FRB formation channels, such as observations of nearby star-forming galaxies\cite{2023A&A...674A.223P} or galaxy clusters\cite{2019MNRAS.490....1A}, would need to account for the noted inclination-related bias.

Furthermore, the observed dependence of the FRB rate on $i$ of the hosts suggests that the majority of FRB sources do not constitute a spherical or halo population, such as globular clusters in the halos of their hosts. If this were the case, one would expect to observe either more FRBs in inclined host bins (Bin 4 in Figure \ref{fig:boxplots}a) than in face-on ones (Bin 1) as a significant fraction of globular clusters are present in the inner halo close to their host disk\cite{2004MNRAS.355..504M}, or no significant difference in FRB distribution if FRB sources are uniformly distributed in their host halo. Therefore, this suggests a disk origin for the majority of FRB sources. Consequently, FRB 20200120E, the closest extragalactic FRB localised to a globular cluster in the Messier 81 galaxy\cite{Bhardwaj+2021a,Kirsten2022Natur}, may be considered as an outlier in the known FRB sample.

While the current sample size of FRB hosts may not definitively explain the observed inclination-related bias, one plausible explanation could involve propagation-induced transmission effects, such as dispersion and scattering. For instance, the contribution of the host galaxy's ISM to the FRB DM likely varies significantly with the inclination angle of the host disk\cite{Xu2015RAA}. In Figure \ref{fig:boxplots}b, we show a box-plot illustrating the range of DM values contributed by FRB hosts (DM$_{\rm host}$) in each of the four bins. The DM$_{\rm host}$ for our sample FRB hosts are listed in Table \ref{tab:host_universe_frbs}. In general, higher DM$_{\rm host}$ values are expected in bins corresponding to higher $i$ (Bins 3 and 4). However, the sample utilised in this study does not exhibit this trend, indicating that we may be overlooking FRBs with significant DM$_{\rm host}$ in Bins 3 and 4. To robustly quantify the impact of DM$_{\rm host}$ on FRB hosts in Bins 3 and 4, we require a larger number of FRB hosts across all four bins. 

The potential scarcity of FRBs with large DM$_{\rm host}$ in Bins 3 and 4 can be linked to pronounced scattering in the host ISM (Methods), akin to observations seen for pulsars within the Milky Way\cite{2016arXiv160505890C}. Scattering has also been previously recognised as a major factor influencing the observed properties and detectability of FRBs\cite{Chawla2022ApJ,Ocker2022ApJ}. Furthermore, both DM$_{\rm host}$ and scattering are expected to be more prominent in star-forming galaxies, where the ISM likely contains a substantial amount of dense, ionised, and turbulent gas\cite{2007ApJ...660..276W}. This can lead to a higher expected scattering (based on the observed star-formation rates (SFR)$-$gas velocity dispersion relation\cite{2022MNRAS.514..480E}) for the same DM compared to a galaxy like our own Milky Way\cite{Salim2014SerAJ}. Additionally, the probability of the FRB sight-line embedded in the Galactic disk intersecting ionised clouds, such as star-forming regions and HII regions, would also be higher in star-forming galaxies. These effects are particularly severe for FRB sources located well within the disk of edge-on galaxies, resulting in a larger inclination-based bias, akin to what is proposed in the case of core-collapse supernovae\cite{1991ARA&A..29..363V}. 

To examine this hypothesis, we analyse the distribution of SFR among FRB host galaxies across the four $i$ bins. Figure \ref{fig:boxplots}c illustrates this distribution for 23 FRB host galaxies, while Figure \ref{fig:boxplots}d focuses on star-forming host galaxies (defined as those with specific SFR (sSFR) $> 10^{-10}$ yr$^{-1}$\cite{2004MNRAS.351.1151B}). Two observations can be made from these plots: Firstly, FRB hosts with higher $i$ exhibit lower median SFRs, indicating a potential inclination-related bias. Secondly, the proportion of star-forming galaxies increases in $i$ Bins 3 and 4. This observation is noteworthy as it hints at the possibility of FRB sources having a preference for star-forming galaxies, despite the anticipated propagation effects being more pronounced in such environments. Such findings lend support to the hypothesis of young FRB progenitor channels\cite{2020ApJ...899L...6L}. However, the limited sample size prohibits definitive conclusions from this analysis. Nevertheless, if this trend persists in future studies, it could elucidate the significant portion of highly scattered FRBs within the CHIME/FRB sample, potentially associated more often with star-forming galaxies\cite{catalog12021ApJS}. 
Unfortunately, we lack scattering parameters (e.g., scattering timescale, decorrelation bandwidth) for most FRBs in our sample to test this prediction.

We also examine the distribution of redshifts among the FRB hosts in our sample across the four $i$ bins, shown in Figure \ref{fig:boxplots}e. We find that all but Bin 4 is consistent with a uniform distribution of FRB redshifts among the four bins, albeit with a small sample size. Given the potential correlation between FRB host properties, such as SFR, and $i$, any bias in the FRB redshift distribution due to host inclinations could potentially bias studies aiming to constrain the origins of FRBs, such as investigations into the evolution of FRB properties relative to SFR or stellar mass. Therefore, if the noted trend is found to be true in case of larger sample of FRB hosts, it would suggest the necessity of correcting for the inclination-related selection effect in the aforementioned analyses.  
Note that $i$ can also influence host demographics studies involving projected FRB offset measurements\cite{Li2020ApJ}, which will be explored elsewhere.

In conclusion, our study highlights the importance of investigating inclination-related biases using a larger sample of both repeating and non-repeating FRBs from upcoming wide-sky sensitive surveys. To mitigate these biases, we recommend focusing on FRB hosts with $i\lesssim 60^\circ$ in future studies. 
\clearpage





\begin{threeparttable}
\begin{center}
\caption{Sample of 23 FRB host galaxies employed in this work}
\label{tab:host_universe_frbs}
\scriptsize
\begin{tabular}{ccccccccccc}\toprule
\textbf{Number}$^{a}$ & \textbf{FRB} & \textbf{Inclination angle}$^{b}$ & 
\textbf{Bin \#} &
\textbf{SFR$_{\rm 0-100 Myr}$}$^{c}$ & \textbf{log(sSFR)} & \textbf{Mass} & \textbf{z} & \textbf{DM$_{\text{host}}$} & \textbf{Host m$_{\rm r}$}$^{d}$ & \textbf{Reference} \\
 & & Degrees  & & M$_{\odot}$ yr$^{-1}$ & log(yr$^{-1}$) & M$_{\odot} \times 10^{10}$ &  & pc cm$^{-3}$ & AB mag&\\
\midrule
1 & 20200906A & 63, 59 & Bin3 & 4.93 & $-$9.43 & 1.33 & 0.3688  & 309 & 19.95 & \cite{Gordon2023ApJ} \\
2 & 20210410D & 41, 41 & Bin1 & 0.03 & $-$10.99 & 0.30 & 0.1415&  468 & 20.65 &  \cite{Gordon2023ApJ} \\
\bf{3} & 20211127I &  21, 26 & Bin1 &35.83 & $-$7.93 & 0.30 & 0.0469& 171 & 14.97 & \cite{Gordon2023ApJ}\\
4 & 20211203C&  7, 11 & Bin1 &15.91 & $-$8.56 & 0.58 & 0.3437 &  395 & 19.64 & \cite{Gordon2023ApJ} \\
5 & 20211212A &  33, 35 & Bin1 &0.73 & $-$10.42 & 1.91  & 0.0707&  127& 16.44 & \cite{Gordon2023ApJ} \\
6 & 20220310F &  39, 41 & Bin1 &4.25 & $-$9.34 & 0.93 &0.4780 & 22 & 20.72 & \cite{Law2023} \\
\bf{7} & 20220914A&  45, 46& Bin2 &0.72 & $-$9.62 & 0.30 &0.1139& 538& 19.77 & \cite{Law2023} \\
8 & 20220920A &  27, 29 & Bin1 &4.1 & $-$9.2 & 0.65 & 0.1582& 170 & 18.16 & \cite{Law2023}  \\
9 & 20221012A &  45, 46 & Bin2 &0.15 & $-$11.81 & 9.77 & 0.2847&   190 & 19.14 &\cite{Law2023} \\
\bf{10} & 20171020A &  69, 65 & Bin3 &0.09 & $-$9.69 & 0.04 & 0.0087 & 83 & 14.97 & \cite{Lee-Waddell2023PASA} \\
\bf{11} & 20190714A &  51, 50& Bin2 &1.89 & $-$9.94 & 1.66 & 0.2365&  335& 20.34 &\cite{Gordon2023ApJ} \\
\bf{12} & 20220207C &  80, 74 & Bin4 &1.16 & $-$9.85 & 0.81 &0.0430 & 149& 18.24 & \cite{Law2023} \\
\bf{13} & 20220509G &  60, 57 & Bin3 &0.23 & $-$11.43 & 6.17 &0.0894 &  154& 17.73 &\cite{Law2023} \\
\bf{14} & 20220825A &  46, 46 & Bin2 &1.62 & $-$9.74 & 0.89 &0.2414&  445& 20.03 &\cite{Law2023} \\
\bf{15} & 20220912A&  43, 46 & Bin2 &2.5 & $-$9.6 & 1.0 &0.077& 34& 19.65 &\cite{2023ApJ...949L...3R} \\
\bf{16} & 20220307B & 44, 47 & Bin2 &2.71 & $-$9.61 & 1.1 &0.2481 & 126& 19.93 &\cite{Law2023}  \\
 17 & 20180924B &  34, 36 & Bin1 &0.62 & $-$10.6 & 2.45 & 0.3212 & 80& 20.33 &\cite{Gordon2023ApJ} \\
18 & 20190102C & 50, 50 & Bin2 &0.4 & $-$10.09 & 0.49 &0.2909 &  94& 20.77 &\cite{Gordon2023ApJ}\\
19 & 20190608B  & 37, 36 & Bin1 &7.03 & $-$9.71 & 3.63  & 0.1178& 238& 17.41 &\cite{Gordon2023ApJ} \\
20 & 20191001A & 53, 56& Bin2 &18.28 & $-$9.47 & 5.37  & 0.2342& 340& 18.36 &\cite{Gordon2023ApJ}\\
\bf{21} & 20220319D & 25, 37 & Bin1 & 0.21 & $-$10.84 & 1.45 & 0.0112&  16& 15.39 &\cite{Law2023} \\
\bf{22} & 20210807D &  41, 41 & Bin1 &0.63 & $-$11.17 & 9.33 & 0.1293 & 66& 17.17 &\cite{Gordon2023ApJ}\\
\bf{23} & 20210320C  & 28, 27 & Bin1 &3.51 & $-$9.82 & 2.34 &0.2796 & 148& 19.47 &\cite{Gordon2023ApJ}\\
    \bottomrule
\end{tabular}

\begin{tablenotes}
\item{$^{a}$: Boldface indicates FRB hosts for which Pan-STARRS DR2 r-band images were used. For all other hosts, we use DESI DR10 r-band images.}
\item{$^{b}$: The first inclination angle value in each row is computed using parametric profile fitting, and the second one is computed with the elliptical isophote fitting. The 1$\sigma$ error in both the estimates is $\approx 1^{\circ}$.} 
\item{$^{c}$: SFR$_{\rm 0\text{\textendash}100 Myr}$ is the integrated 0\text{\textendash}100 Myr star-formation rate taken from the references stated in the last column.} 
\item{$^{d}$:These values are estimated after subtracting the fiducial contributions of the Milky Way ISM\cite{NE20012002}
, and intergalactic medium (using the median value of the Macquart relation\cite{Macquart2020Nature}) from the FRB DMs, which are reported in their respective discovery papers.}
\end{tablenotes}

\end{center}
\end{threeparttable}

\begin{table}
\begin{center}
\caption{Inclination angle bin range derived using mean SDSS galaxy $\cos{i}$ CDF}
\label{tab:bin_range}
\begin{tabular}{cc}
\hline
\textbf{Name} & \textbf{Inclination angle range}$^{a}$\\
& Degrees \\
\hline
Bin 1 & [0 $-$ 42)\\
Bin 2 & [42 $-$ 60)\\
Bin 3 & [60 $-$ 76)\\
Bin 4 & [76 $-$ 90]\\
\hline
\end{tabular}

\begin{tablenotes}
\small {
\item{$^{a}$: Parentheses `)' indicate that the endpoint is not included in the range, while square brackets `]' indicate that the endpoint is included.}
}
\end{tablenotes}
\end{center}
\end{table}


\clearpage

\textbf{\Large\bf References:} \\

\bigskip

\bibliographystyle{naturemag}
\bibliography{references}


\newpage

\section*{\LARGE Methods}
\label{sec:methods}


\bigskip
\bigskip

\subsection{Inclination Angle Estimation:}
\label{sec:inclination_angle}
In this work, we define $i$ as the angle between the normal vector to the galactic disk and the observer's line of sight. 
Throughout this work, we estimate $i$ using axial-ratio $b/a$ and the Hubble formula\citemethods{1926ApJ....64..321H} which works well for most of the spiral galaxies,

\begin{center}
    \begin{equation}
  i = \text{acos}\biggl( \sqrt{\frac{{(b/a)}^{2} - q_{0}^{2}}{1.0 -q_{0}^{2}}}~\biggr),
\label{eqn:hubble}
\end{equation}

\end{center}
where $q_{0}$ is the axial-ratio for an edge-on system, which is assumed to be 0.2\citemethods{1988PASP..100..344V,Unterborn2008ApJ}. Note that this formula gives unreliable $i$ measurements particularly at $b/a < 0.2$. It is therefore suggested to discard those galaxies from the analyses that employ Equation \ref{eqn:hubble}\citemethods{Japelj2018A&A}. Fortunately, none of our FRB host galaxies meet this criterion, thus ensuring no impact on our analysis.

The inclination angle of disk galaxies is often determined by assuming that the disk is circular when viewed face-on, possesses axis symmetry, and has a finite thickness (quantified by the parameter q$_{0}$ in Equation \ref{eqn:hubble}). However, it is important to acknowledge that the assumption of axis-symmetric circular disks when viewed face-on is often violated. Nonetheless, this does not have any meaningful impact on the results presented in this study, as discussed in the next section.



\bigskip
To determine a galaxy model that effectively represents the morphology of FRB host galaxies, we employ two methods: (1) parametric profile fitting and (2) elliptical isophote fitting.

We perform parametric profile fitting of FRB hosts using \texttt{AutoProf}, an automated Python-based non-parametric light profile extraction package\cite{Stone2021b}. After extracting the surface brightness profile from the deepest available archival r-band images of the host (DESI or Pan-STARRS if DESI does not cover the FRB field-of-view), \texttt{AutoProf} offers modules for characterising galactic structures by fitting the light distribution to several predetermined profiles. Our primary choice is the S\'ersic profile, which is a generalised fitting function which is widely used in structural studies of galaxies\citemethods{1963BAAA....6...41S}. However, for face-on spiral hosts that exhibit dominant spiral arms and bars, where the S\'ersic profile may be inadequate, we use alternative models such as Superelliptical, Gaussian or Spline profiles (used for only three sources, FRB Hosts 12, 15, and 22 in Table \ref{tab:host_universe_frbs}, respectively), as recommended by the \texttt{AutoProf} authors\cite{Stone2021b}. Furthermore, we evaluate the quality of the model fitting by analysing the model convergence value, which is provided as one of the fitting outputs. We set a maximum threshold for that to be $5 \times 10^{-5}$ for successful fitting, in accordance with the package recommendation. Figure 4 displays the fitted model profile and residual images alongside the r-band imaging data used in this analysis. It is noteworthy that in cases where the S\'ersic profile yields satisfactory fitting results, the Sérsic index is found to be $\leq$ 1, which suggests that the FRB hosts in our sample are disk-dominated. Furthermore, \texttt{AutoProf} provides the axial-ratio estimate ($b/a$) of the best-fitted model, which we use in Equation \ref{eqn:hubble} to calculate the inclination angle ($i$) of the FRB hosts. The estimated $i$ values are presented in Table \ref{tab:host_universe_frbs}. 

To ensure that the $i$ values derived from the parametric profile fitting method are not biased by irregularities in the FRB host galaxy structure, we also estimate $i$ using the elliptical isophote fitting method, which models galaxies without assuming any specific profile\citemethods{Jedrzejewsk1987MNRAS}. For this purpose, we utilise \texttt{Photutils}, an Astropy-affiliated open-source Python package that offers tools for detecting and performing photometry of astronomical sources\citemethods{larry_bradley_2023_7946442}. \texttt{Photutils} provides modules that fit elliptical isophotes as contours of constant surface brightness to a galaxy image. It captures the galaxy's projected two-dimensional light distribution by iteratively adjusting the geometrical parameters of a trial ellipse. To estimate morphological parameters of the FRB hosts, such as ellipticity ($\epsilon$), we utilise the isophote that encompasses 95\% of the galaxy's light. This approach provides a robust geometric proxy for gauging the three-dimensional orientation of the galaxies under study\cite{Ciambur2015ApJ}. The best-fitted profile and residual images of the FRB hosts obtained from this method are displayed in Figure 5. We then calculate $i$, using the Hubble formula (Equation \ref{eqn:hubble}), where the axial-ratio ($b/a$) values are estimated using the relation, $b/a  = 1-\epsilon$\cite{larry_bradley_2023_7946442}. The $i$ values estimated using the isophote elliptical fitting are shown in Table \ref{tab:host_universe_frbs}.

Note that the 1$\sigma$ uncertainties in $i$ values obtained from both parametric profile fitting and isophote elliptical fitting methods are $\approx$ $1^{\circ}$, which are computed from the output covariance matrix provided by both the packages. Upon comparing the two $i$ estimates (see Table \ref{tab:host_universe_frbs}), we observe that both methods yield similar $i$ estimates, with a few outliers (for example, FRBs 20220207C and 20220319D). This discrepancy is not unexpected, given the distinct underlying assumptions of the methodologies employed by the two methods. Therefore, to ensure a fair comparison, we utilise \texttt{AutoProf} $i$ estimates when comparing them with those derived from SDSS galaxies, which are determined using a similar parametric profile fitting method. Importantly, we want to emphasise that the results of this study remain unchanged even if we were to utilise \texttt{Photutils} $i$ estimates. This underscores the insignificance of the observed difference in $i$ values to our overall conclusions.
Furthermore, we estimate the root mean square difference (RMSD) between the two samples to be 3.7$^{\circ}$. This difference could be considered as a potential systematic error in our $i$ estimates that is not captured by the covariance matrix. Based on this, only one FRB, FRB 20220319D, exhibits an absolute $i$ difference exceeding 2 times the RMSD. If we exclude this FRB host from our analysis and consider the remaining 22 FRB hosts in our cumulative distribution function (CDF) analysis, we still find that the cos($i$) distributions of the FRB host and SDSS galaxy samples are statistically inconsistent, with an AD test p-value $< 0.001$. Therefore, we argue that our conclusions remain robust despite observed differences in the $i$ estimates obtained from the two methods.

\bigskip
\subsection{Inclination angle estimation of randomly sampled SDSS field galaxies:}

As noted above, the accuracy of the Hubble formula is known to vary based on the physical properties of galaxies. This is largely due to the presence of non-axisymmetric structures, such as bars and spiral arms, and/or strong central bulges, which are present in most disk galaxies, or other stellar population effects, such as uneven dust extinction on the near and far halves of galaxies, which hinder accurate measurements of disk $i$. Moreover, the assumption of a universal q$_0$ is likely incorrect, as the thickness of a disk is not constant across different galaxy types and masses. All these effects have the potential to introduce bias into our analysis if we simply compare estimates of $\cos(i)$ for FRB host galaxies from our analysis with an expected uniform distribution based on the assumption of random orientation of galaxies in the Universe. To minimise the impact of these biases, we adopt a methodology where the CDF of $\cos(i)$ for FRB hosts is compared with that of randomly selected SDSS field galaxies, where the $i$ estimates face similar biases.To demonstrate this, we present in Figure \ref{fig:sdss_hist_axis_symmetric} the probability distribution function of the cosine of the observed $i$ of a random sample of 10,000 SDSS galaxies, which are estimated based on the methodology described below. This plot clearly indicates that the SDSS $i$ is also affected by the aforementioned biases. Therefore, we argue that the potential inaccuracies in the $i$ estimates derived from the Hubble formula are unlikely to substantially impact the principal conclusions of our study.

To identify any inclination angle-based selection bias in the FRB host sample, we compare the CDF of $\cos(i)$ of the FRB host sample with that of randomly selected disk-dominated late-type galaxies. For this comparison, we utilise the SDSS DR16 catalogue\cite{SDSSDR16_2020} and randomly sample 23 galaxies. To ensure that our SDSS sample closely aligns with our FRB host sample, we select galaxies from the SDSS DR16 \texttt{PhotoObjAll} table that satisfy the following criteria:

\begin{enumerate}
\item mode = 1: Classified as a primary survey object.
\item clean = 1: Good quality photometry exists.
\item class = 3: Classified as a galaxy.
\item lnLExp\_r $>$ lnLDeV\_r: Identifies galaxies where the exponential disk profile fits better than the de Vaucouleurs fit (to select disk-dominated galaxies).
\item $u - r < 2.3$: Criterion to identify late-type galaxies in the SDSS data\cite{Strateva2001AJ}.
\item m$_{\rm r}<$ 21 mag: Same constraint used to select FRB hosts in our study.
\end{enumerate}

Note that criteria 1 and 2 are intended to remove possible artifacts or duplicate objects in our sampling, while criteria 3, 4, and 5 are utilised to identify disk-dominated late-type galaxies. Finally, criterion 6 is employed to apply the same constraint used to select FRB hosts in this study. To mitigate the potential impact of a small sample size, we generate 10,000 realisations of the data sampling. Using these realisations, we estimate the 68\% uncertainty region on the cos($i$) CDF of SDSS galaxies, depicted as a blue-shaded region in Figure \ref{fig:main_cdf}. As depicted in Figure \ref{fig:main_cdf}, the two cumulative distribution functions (CDFs) are observed to represent statistically distinct distributions (p-value $= 0.0003$ using the AD test from \texttt{Scipy}).

Next, to ensure that the observed differences in the CDFs between our FRB host galaxy samples and randomly sampled SDSS galaxy samples are not influenced by the criteria applied to select late-type SDSS galaxies, we analyse the impact of Criteria  5 and 6 on the aforementioned result.

\bigskip

\subsection{Colour constraint}
To assess whether our results are sensitive to different u$-$r colour thresholds, we repeat the SDSS galaxy sampling under the following conditions: u$-$r $< 1.5$,  u$-$r $< 2.3$, u$-$r $<$ 3, and u$-$r $<$ 3.5 (likely early-type). The CDFs of the randomly sampled SDSS galaxies for each of these colour thresholds are shown in Figure \ref{fig:sdss_biases}(top). From this analysis, we conclude that the u$-$r colour constraint employed to select SDSS galaxies as FRB hosts does not impact the conclusions of our work.

\bigskip

\subsection{m$_{\rm r}$ magnitude constraint}
Figure \ref{fig:sdss_biases} (bottom left) illustrates the CDFs of $\cos(i)$ for SDSS galaxies, generated using the same formalism as described above, except that we employ three different m$_{\rm r}$ threshold: m$_{\rm r} < 21$ AB (the original threshold), m$_{\rm r}$ $< 20$ AB, and m$_{\rm r}$ $< 19$ AB. While the CDFs for the three m$_{\rm r}$ thresholds do not completely overlap, the observed CDFs are still statistically inconsistent with that of FRB host galaxies (each yielding a p-value $< 0.0005$ using the AD test from \texttt{Scipy}). To further assess the dependence of our conclusions on m$_{\rm r}$, we randomly sample SDSS galaxies such that the SDSS galaxies in each sampling shows the same distribution of r-band magnitude as seen in the case of our FRB host sample. The resultant CDF of $\cos(i)$ is shown in Figure \ref{fig:sdss_biases} (bottom right). Here too, we conclude that the observed $\cos(i)$ distribution of FRB hosts is inconsistent with what we expect from random sampling of galaxies in the Universe.

{\bf\Large Role of Scattering in the observed inclination-related bias}

Before attributing the observed host inclination-related bias in FRB detections to scattering, it is essential to recognise that different telescopes have unique response functions. To derive meaningful constraints on the astrophysical aspects of any detection pipeline, an absolute calibration of selection effects is necessary to account for systematic biases arising from both the telescope and the software detection pipeline. The CHIME/FRB collaboration demonstrated the significance of this approach by using Monte Carlo-style real-time injections into their software system\cite{catalog12021ApJS,Merryfield2023AJ}. Their analysis revealed significant selection effects for scattering, with fractional completeness varying by orders of magnitude across detected values. Notably, there is a severe selection bias against CHIME/FRB events with scattering times greater than 30 ms at 600 MHz. Unfortunately, such selection functions and related studies have not been conducted for the CRAFT and DSA-110 surveys, preventing a robust analysis of scattering effects on the observed bias.

With this limitation in mind, we proceed to quantitatively validate whether scattering can explain the observed inclination-related bias. In the absence of a completeness limit for the scattering timescale ($\tau_{\rm scatt}$) for the CRAFT and DSA-110 surveys, we scale the completeness limit of scattering timescale $=$ 30 ms estimated by the CHIME/FRB collaboration at 600 MHz to $\approx$ 4 ms at 1 GHz using the following scaling relation: $\tau_{\rm scatt} \sim \nu^{-4}$. At first-order approximation, this scaling is reasonable because the all-sky rates of FRBs detected at the two frequencies and the demographics of their host galaxies are found to be similar \citemethods{Gordon2023ApJ,catalog12021ApJS,2024ApJ...961...99I}, suggesting that the detected FRB population may not vary significantly between these two frequencies. Additionally, we note that the median $\tau_{\rm scatt}$ at 1 GHz, estimated using nine FRBs from our sample where $\tau_{\rm scatt}$ values are available\cite{2019Sci...365..565B,2020MNRAS.497.3335D,2020MNRAS.497.1382Q,2020ApJ...901L..20B,2023MNRAS.524.2064C,2023ApJ...949L..26C,2023ApJ...955..142Z}, is $\approx$ 1 ms. This is within the completeness threshold assumed above. However, we acknowledge that this treatment is far from ideal since all three telescopes and their detection pipelines can have different systematics. However, given the absence of necessary information, this approach is reasonable in our case, as our discussion aims solely to validate the potential of scattering to explain the reported inclination-induced effect.


In a homogeneous and isotropic Universe with no particular preference in the orientation of intervening galaxies, we do not expect to see any inclination dependent bias in our sample if the scattering from foreground galaxies and their halos dominates. Therefore, the observed inclination-related bias in our observation is most likely the result of the scattering in FRB hosts. Moreover, for low-redshift FRBs which constitute our sample (z $<$ 0.5), the contribution of intervening galaxy disk and their halos should not be significant (see Figure 7 of \citemethods{Ocker2022ApJ}). Hence, in our subsequent discussion, we only focus on the scattering in FRB hosts.

We estimate the DM contributed by a scattering layer, DM$_{l}$ (in pc cm$^{-3}$), at redshift $z$, assuming it is located in the host. This estimation aims to produce a scattering timescale, $\tau_{\rm scatt}$, of 4 ms at 1 GHz using Equation 1 from \cite{Ocker2022ApJ}:

\[
\tau(\text{DM}_{l}, z) \approx 48.03 \text{ ns} \, A_{\tau} (1+z)^{-3} \widetilde{F} G_{\rm scatt} \text{DM}_l^2
\]

Here, $A_{\tau} \approx 1$ converts the mean scattering delay to the 1/$e$ time typically estimated from pulse shapes, $G_{\rm scatt}$ is the dimensionless geometric leverage factor, and $\widetilde{F}$ quantifies turbulent density fluctuations in the scattering layer. Assuming the source-to-observer distance is much larger than the scattering medium's thickness, with the source and the scattering layer embedded in the same medium, we set $G_{\rm scatt} = 1$. Additionally, we assume $\widetilde{F} = 1$, as is commonly considered for the Milky Way-like thin disk (see Table 1 of \cite{Ocker2022ApJ}). For a mean redshift of $z \approx 0.07$ estimated using FRB hosts in our sample, we find DM$_{l} \approx 320$ pc cm$^{-3}$. 
This can be readily obtained for FRBs embedded in inclined host galaxies ($> 60^\circ$), as shown in several analytical studies\cite{Xu2015RAA,Ocker2022ApJ}. Moreover, recent work using the Feedback in Realistic Environments (FIRE-2) cosmological zoom-in simulation suite studied the effect of inclination angle on the FRB host DM contribution \cite{2024arXiv240603523O}, where the authors found that for L$_{\star}$ disk galaxies with $i > 60^\circ$, the average DM contribution is $\gtrsim 200$ pc cm$^{-3}$. 

In the above calculation, we ignored the contribution from the FRB circumburst medium, which could be significant, as seen in the case of FRB 20190520B\cite{2023Sci...380..599A}. Additionally, in highly inclined galaxies, the FRB sight-line might intersect a compact scattering structure, such as a plasma lens, within its host galaxy. This could too substantially contribute to $\tau_{\rm scatt}$\cite{2023MNRAS.519..821O}. Both of these factors can play crucial roles in star-forming host galaxies of FRBs, potentially reducing the estimated DM$_{l}$ above. Moreover, including the effects of intervening halos could further reduce this estimate. Therefore, scattering within FRB hosts emerges as a promising explanation for the observed inclination-related bias.

\bigskip



\clearpage

\bigskip

\bibliographystylemethods{naturemag}
\bibliographymethods{references}



\begin{addendum}
	
\item[Author Contributions]

The methodology and framework of this work were conceived by MB. KJ conducted the initial analysis for this project under the supervision of MB. However, all the results presented in this work were refined by JL with the assistance of MB. MB led the writing of this manuscript, with JL contributing to the Methods section.


\item[Acknowledgments]

We are grateful to Dr. Dustin Lang for valuable and insightful discussions. We also thank Dr. Connor Stone for assisting with profile fitting routines of \texttt{AutoProf}. This research made use of Photutils, an Astropy package for detection and photometry of astronomical sources.
 M.B is a McWilliams fellow and International Astronomical Union Gruber fellow. M.B. receives support from the McWilliams seed grant.

\item[Competing Interests]

The authors declare no competing interests.


\item[Correspondence and Requests for Materials]

Correspondence should be addressed to Dr. Mohit Bhardwaj (\href{mailto:mohitb@andrew.cmu.edu}{mohitb@andrew.cmu.edu}).


\item[Data Availability]

The data that support the plots within this paper and other findings of this study are available 
from the corresponding author upon reasonable request. The Pan-STARRS DR1, SDSS DR16, and DESI optical r-band imaging data used in this analysis can be accessed from the publicly available MAST SDSS Data Archive Server, MAST PS1 Science Archive Server, and DESI Legacy Imaging Survey interface, respectively. 

	
\item[Code Availability]

The following software packages were used to analyze the data presented in this paper: \break \texttt{APLpy}\,\citemethods{ForemanMackey+2013}, \texttt{astropy}\,\citemethods{Astropy+2013},
\texttt{AutoProf}\,\cite{Stone2021b}, \texttt{Photutils}\,\citemethods{larry_bradley_2023_7946442}, and
\texttt{Scipy}\,\citemethods{2020SciPy-NMeth}. 
The codes used for data processing and producing the figures are available from 
the corresponding author upon reasonable request.

\end{addendum}



\renewcommand{\figurename}{\textbf{Extended Data Figure}}
\setcounter{figure}{0}


\pagebreak

\begin{figure*}
    \caption{Parametric profile fitting results for 23 FRB host galaxies using the \texttt{AutoProf} package: For each galaxy, we show three images - (left) r-band image of the host, (center) best-fitted galaxy model, and (right) resultant residual. The numbers in the top left of each r-band image correspond to the host number and the estimated inclination angle (in degrees, quoted in parentheses), as shown in Table \ref{tab:host_universe_frbs}. The X-axis and Y-axis of all plots represent right ascension and declination, respectively.}
        \label{fig:autophot}
    \centering
    \begin{minipage}{0.49\textwidth}
        \centering
        \includegraphics[width=\linewidth]{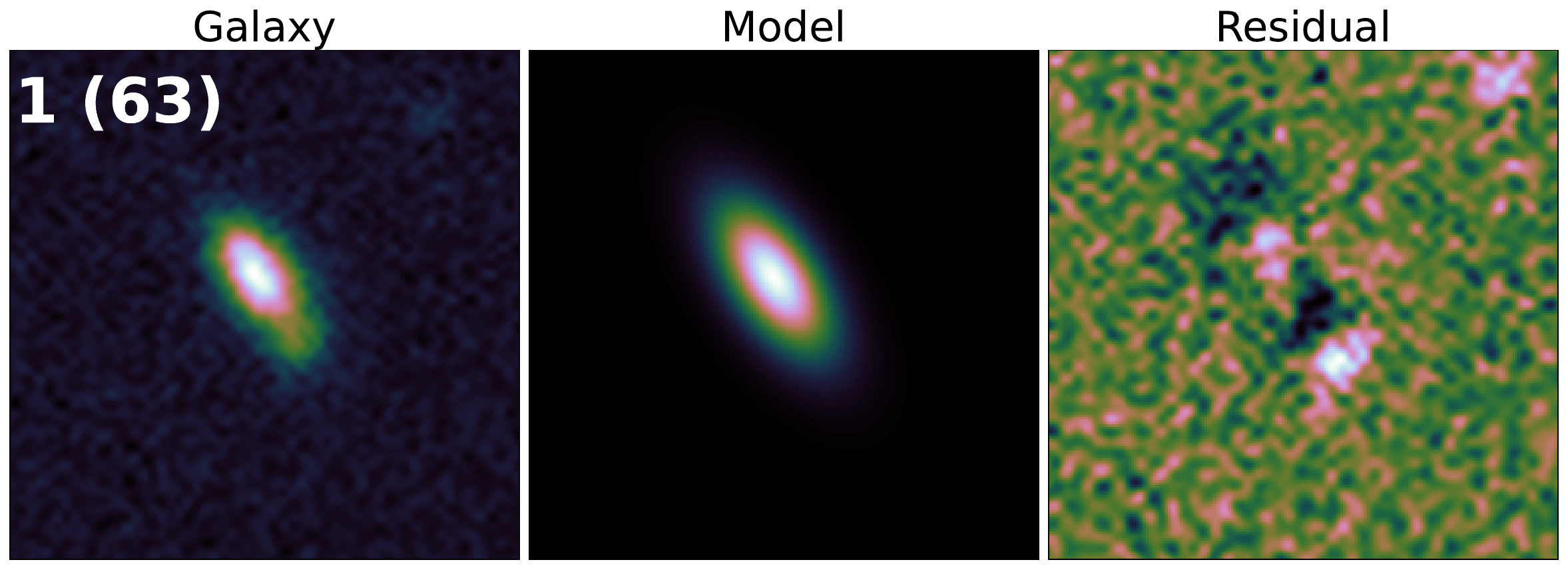}
    \end{minipage}\hfill
    \begin{minipage}{0.49\textwidth}
        \centering
        \includegraphics[width=\linewidth]{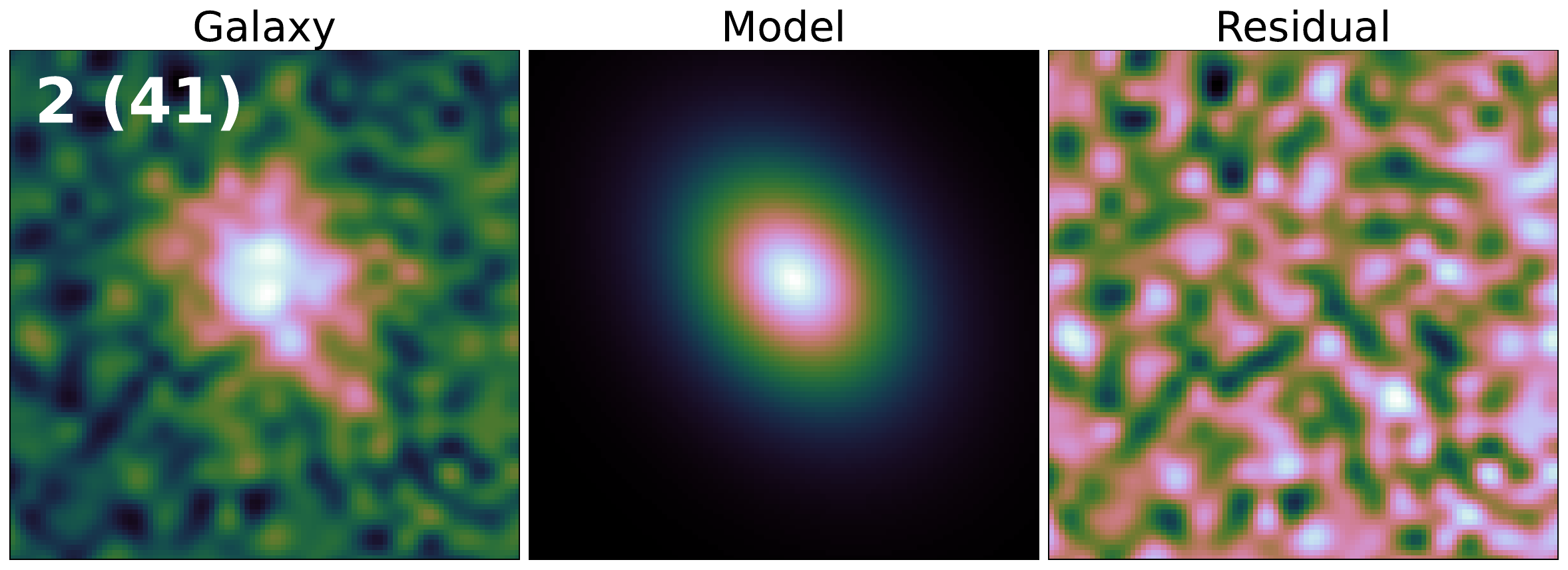}
    \end{minipage}
    
    \vspace{0.5em}
    
    \begin{minipage}{0.49\textwidth}
        \centering
        \includegraphics[width=\linewidth]{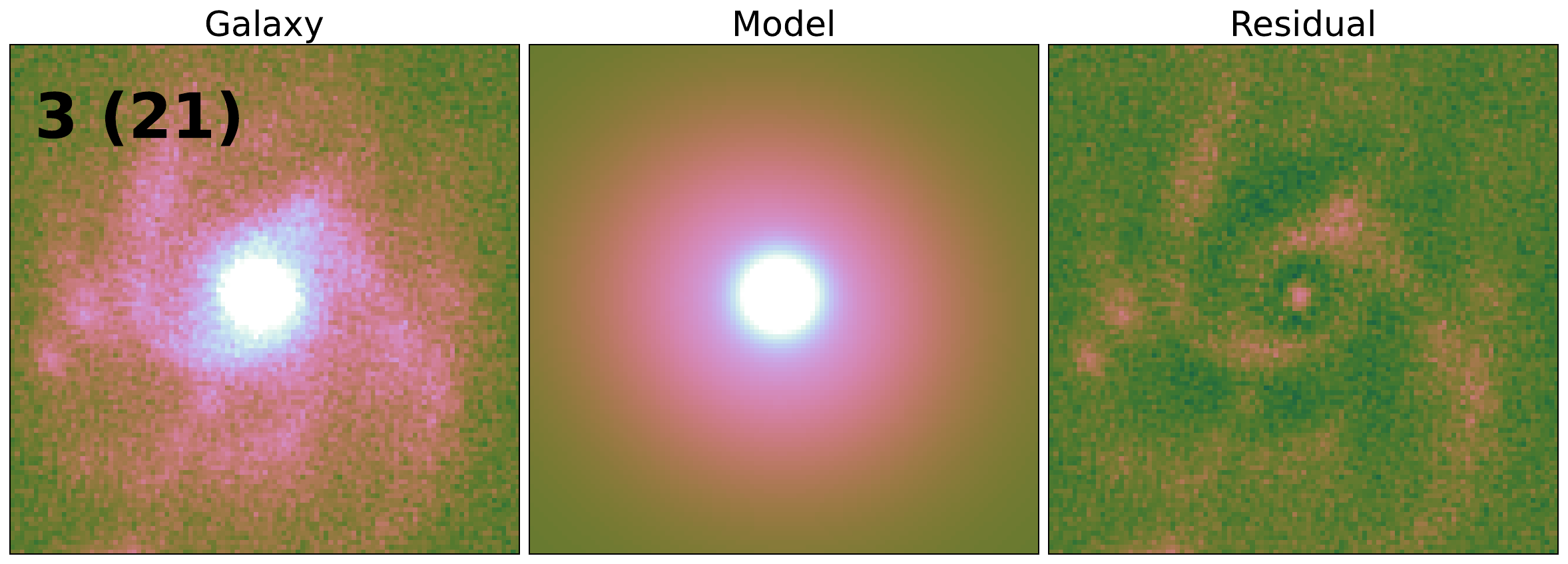}
    \end{minipage}\hfill
    \begin{minipage}{0.49\textwidth}
        \centering
        \includegraphics[width=\linewidth]{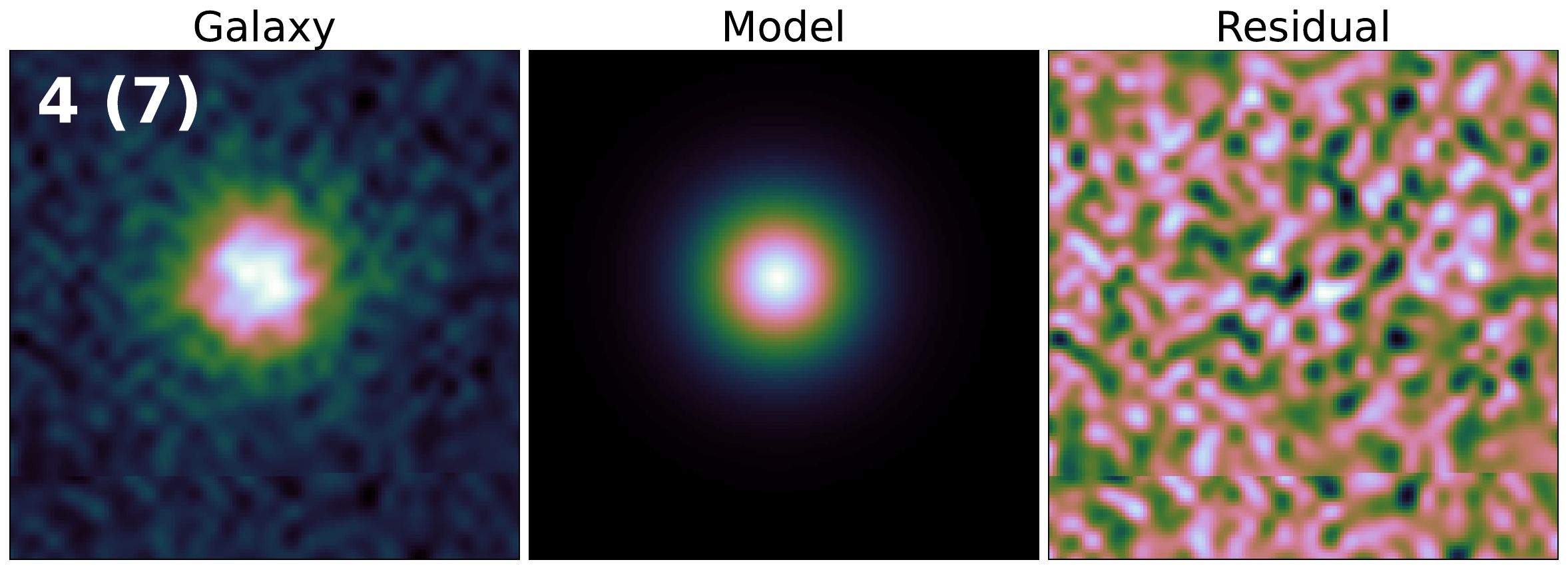}
    \end{minipage}
    
    \vspace{0.5em}
    
        \begin{minipage}{0.49\textwidth}
        \centering
        \includegraphics[width=\linewidth]{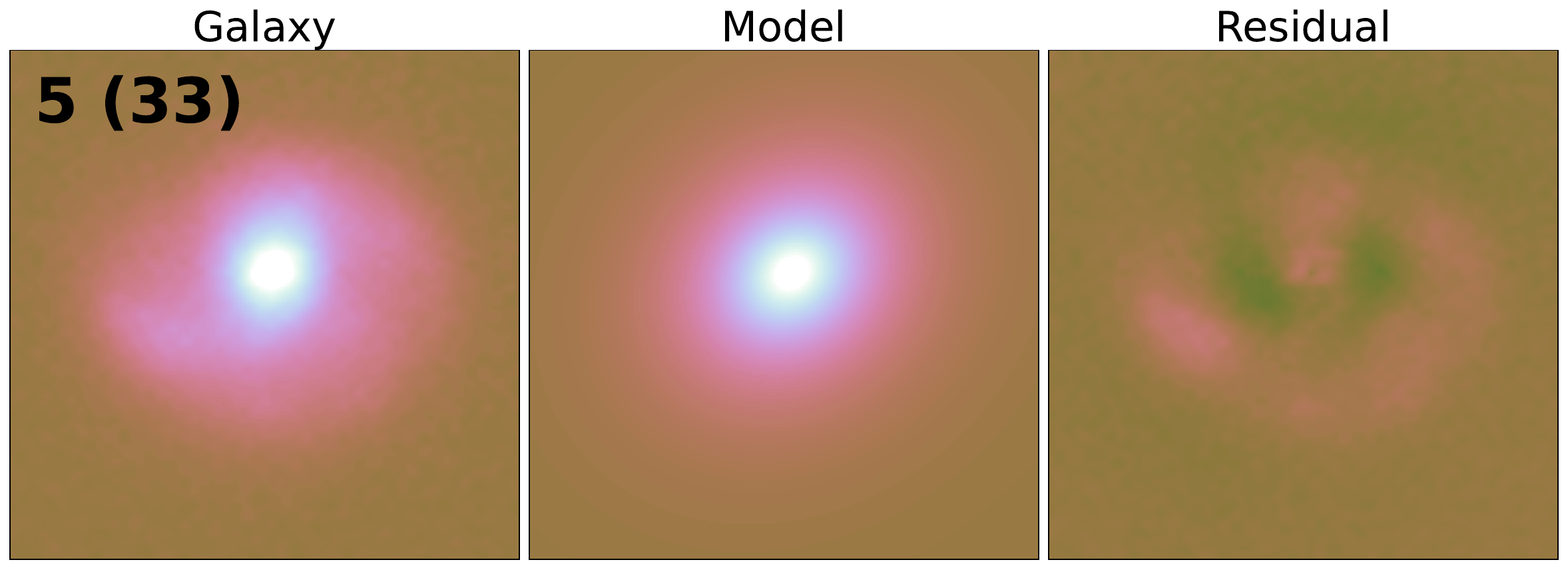}
    \end{minipage}\hfill
    \begin{minipage}{0.49\textwidth}
        \centering
        \includegraphics[width=\linewidth]{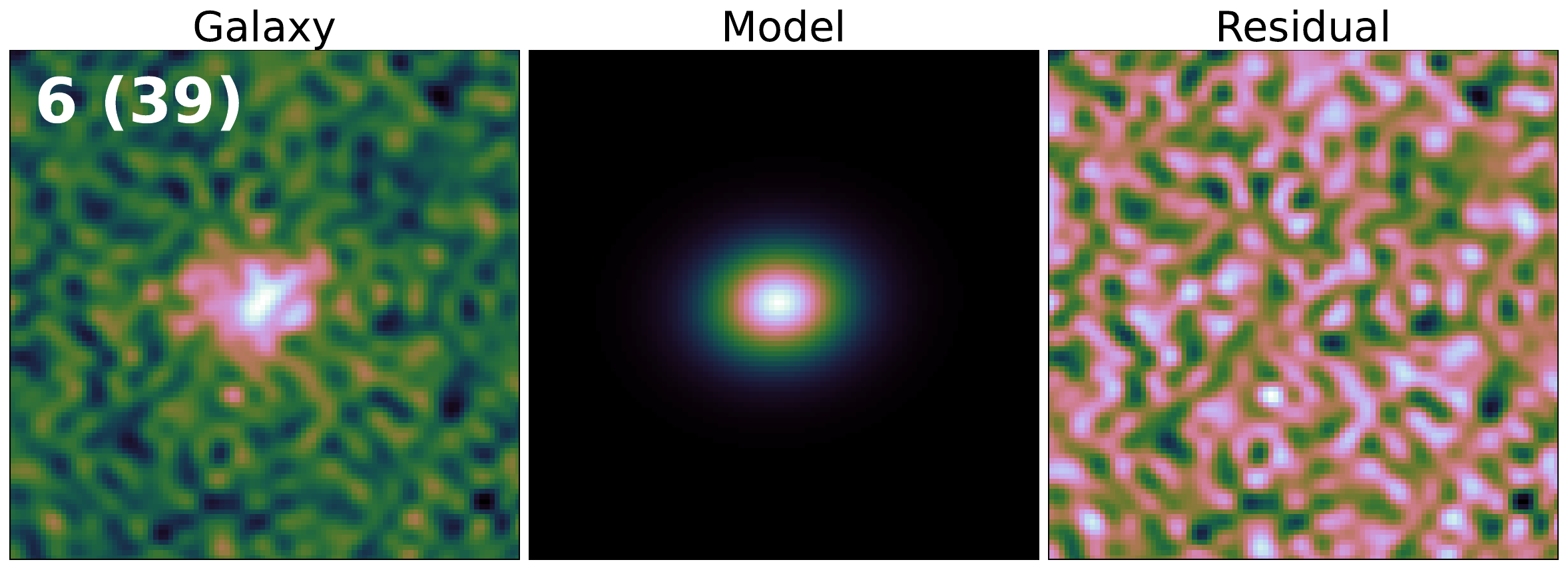}
    \end{minipage}
    
    \vspace{0.5em}
    
        \begin{minipage}{0.49\textwidth}
        \centering
        \includegraphics[width=\linewidth]{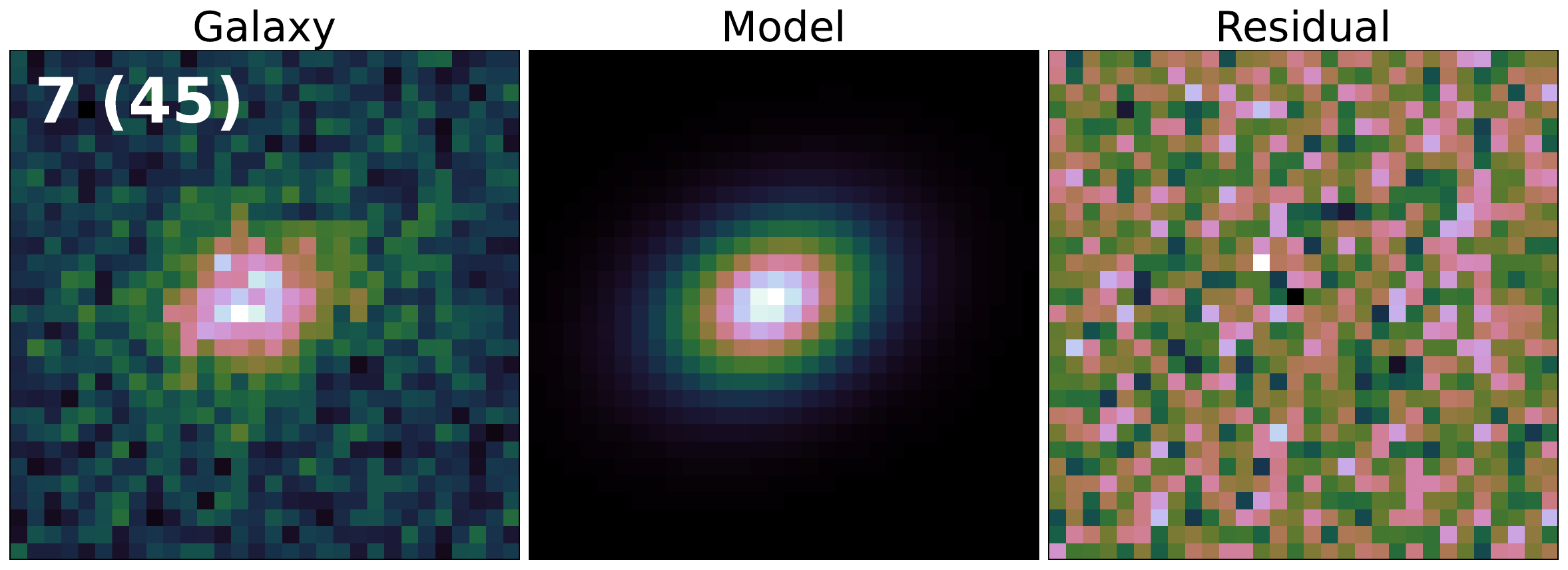}
    \end{minipage}\hfill
    \begin{minipage}{0.49\textwidth}
        \centering
        \includegraphics[width=\linewidth]{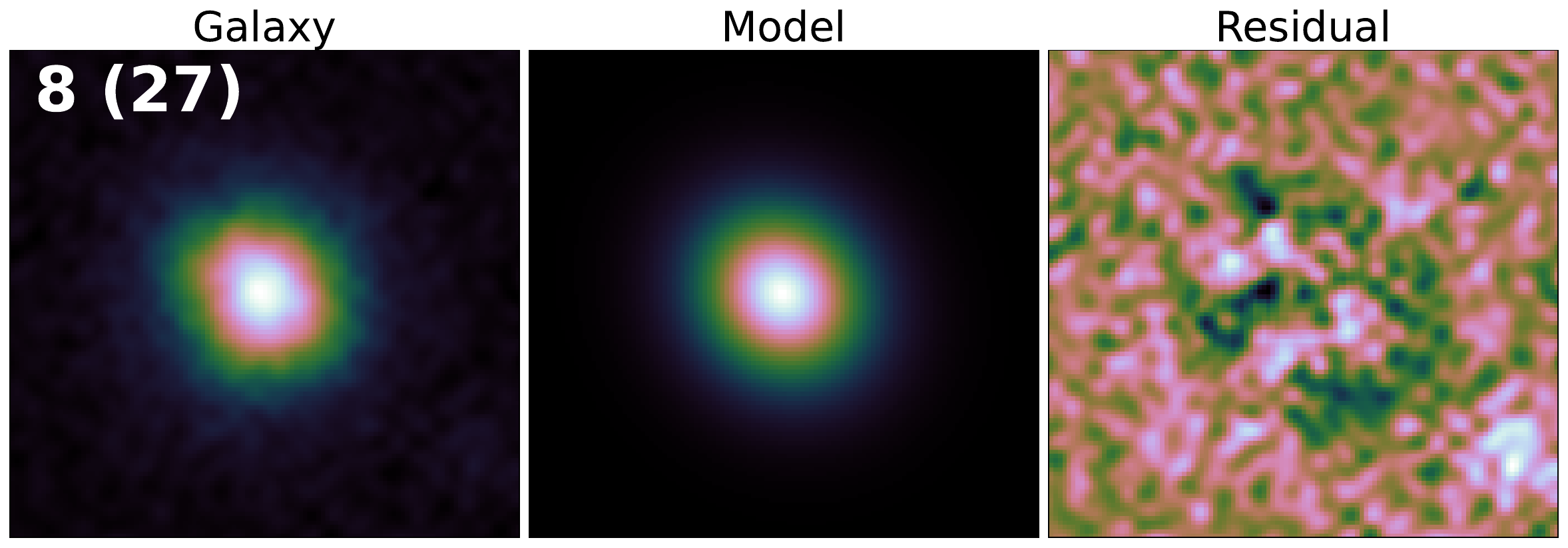}
    \end{minipage}
    
    \vspace{0.5em}
    
    \begin{minipage}{0.49\textwidth}
        \centering
        \includegraphics[width=\linewidth]{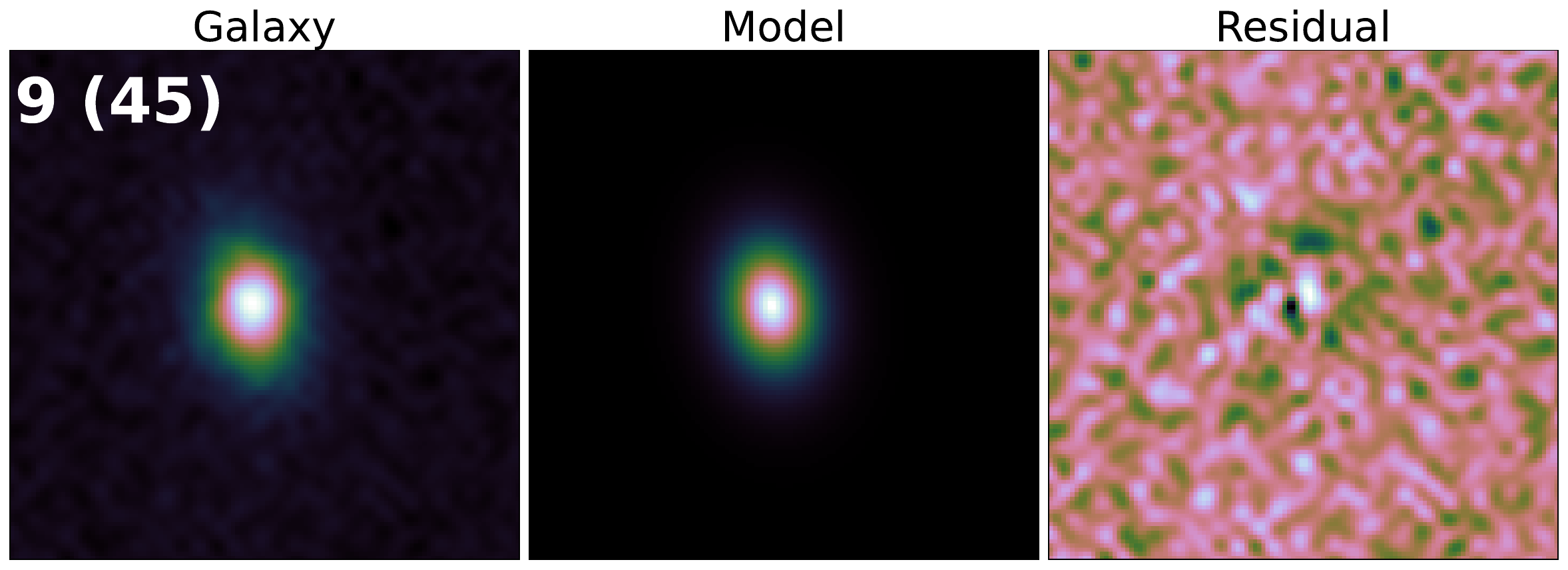}
    \end{minipage}\hfill
    \begin{minipage}{0.49\textwidth}
        \centering
        \includegraphics[width=\linewidth]{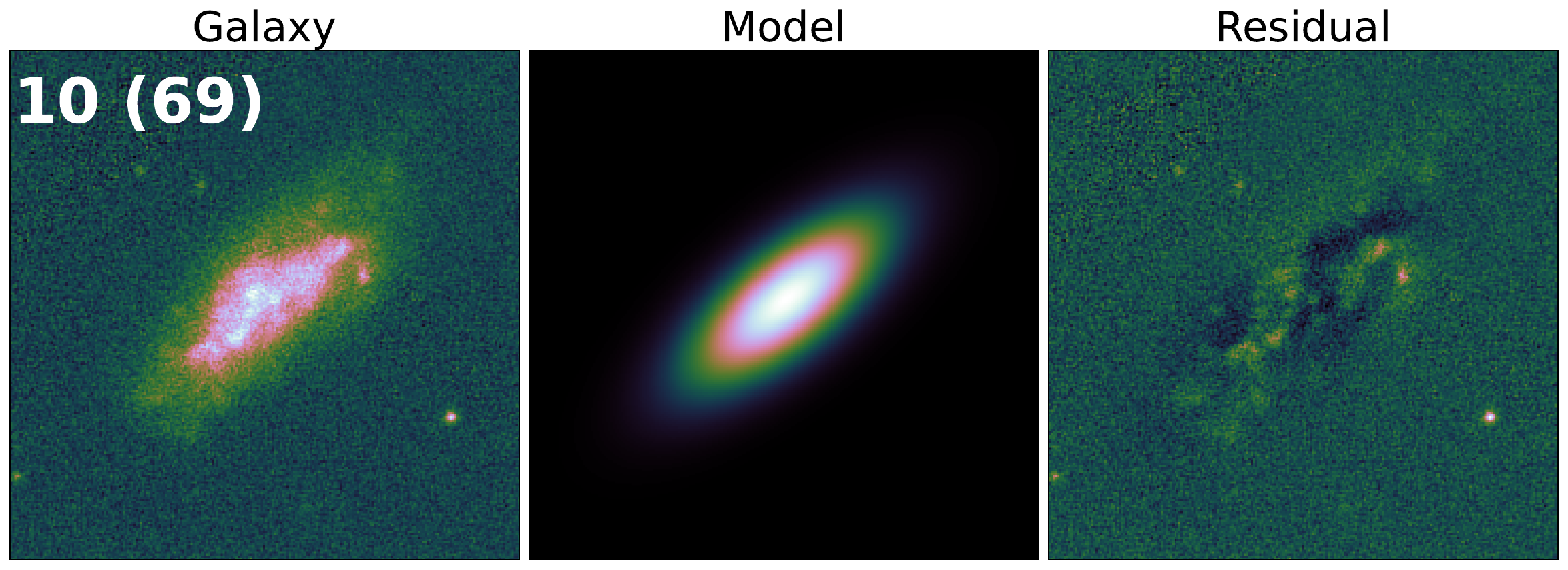}
    \end{minipage}
    
    \vspace{0.5em}
    
        \begin{minipage}{0.49\textwidth}
        \centering
        \includegraphics[width=\linewidth]{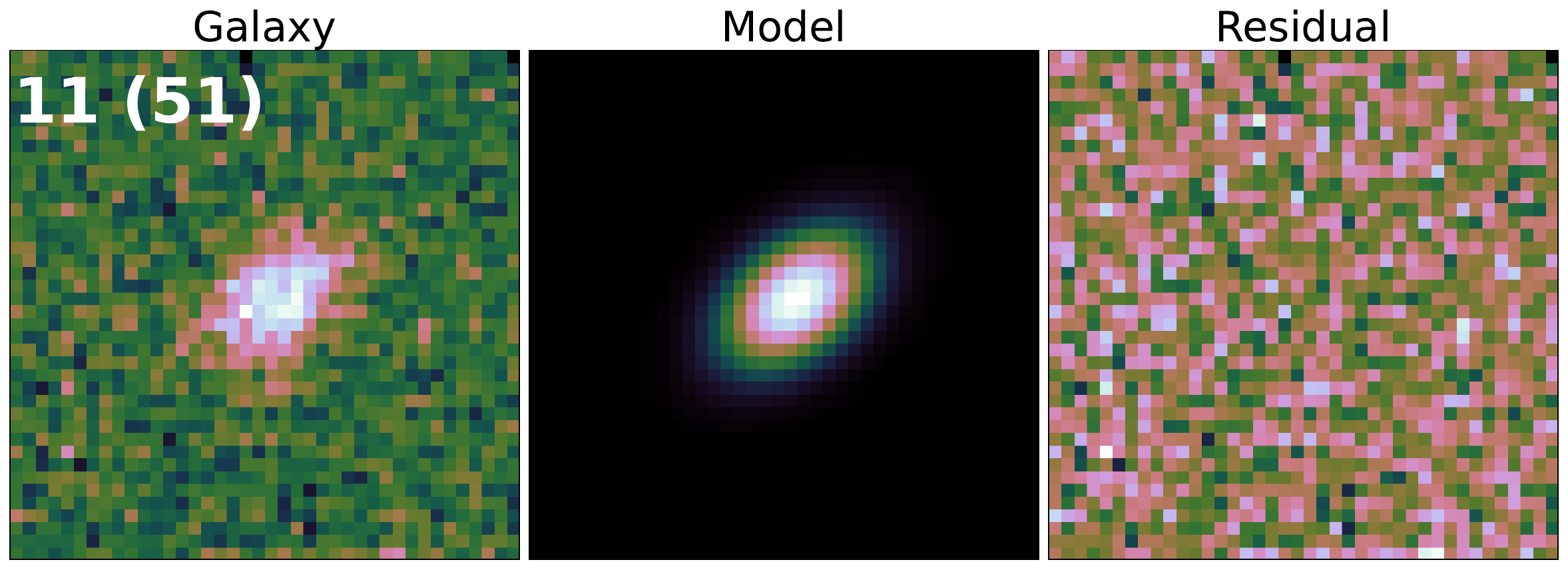}
    \end{minipage}\hfill
    \begin{minipage}{0.49\textwidth}
        \centering
        \includegraphics[width=\linewidth]{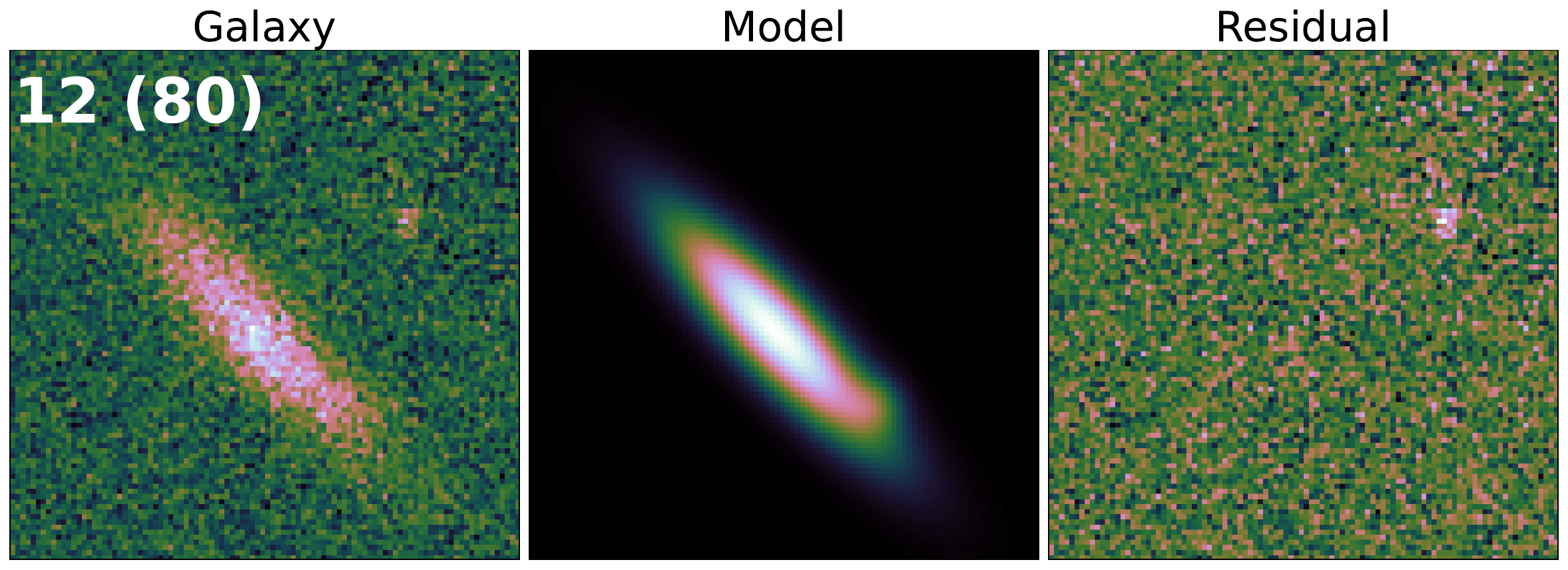}
    \end{minipage}

    \phantomcaption
\end{figure*}

\begin{figure*}
\ContinuedFloat
    
            \begin{minipage}{0.49\textwidth}
        \centering
        \includegraphics[width=\linewidth]{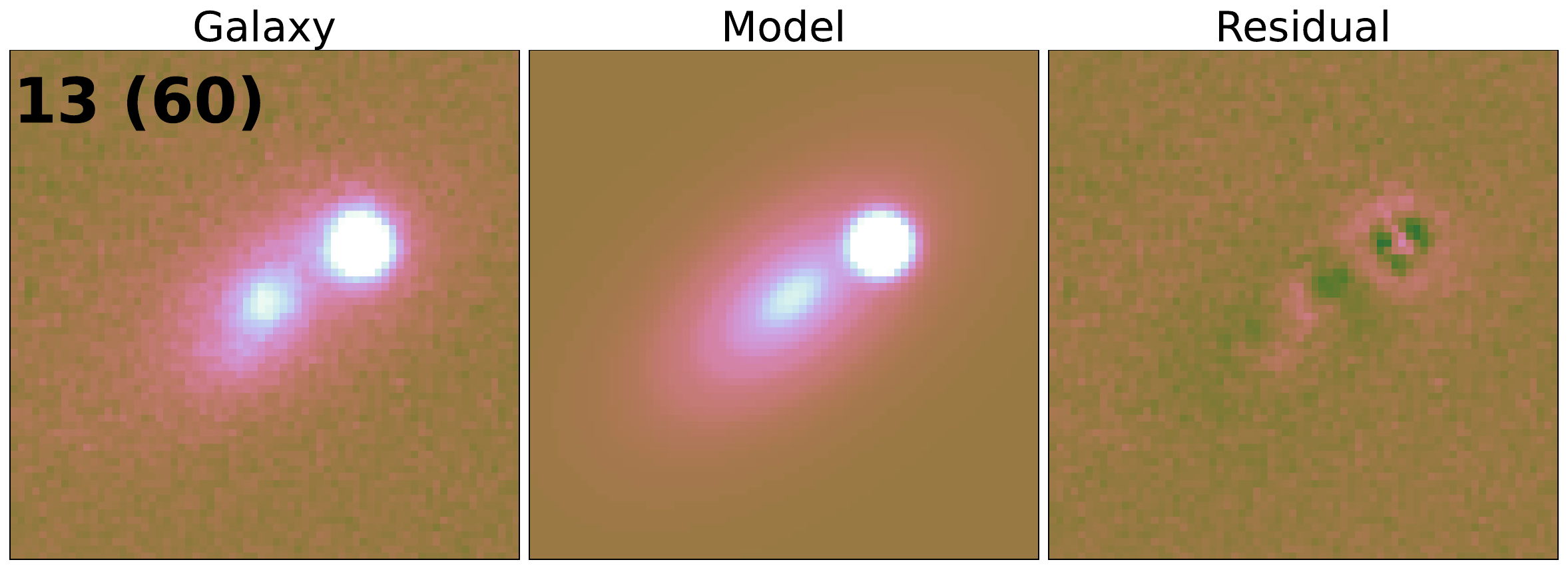}
    \end{minipage}\hfill
    \begin{minipage}{0.49\textwidth}
        \centering
        \includegraphics[width=\linewidth]{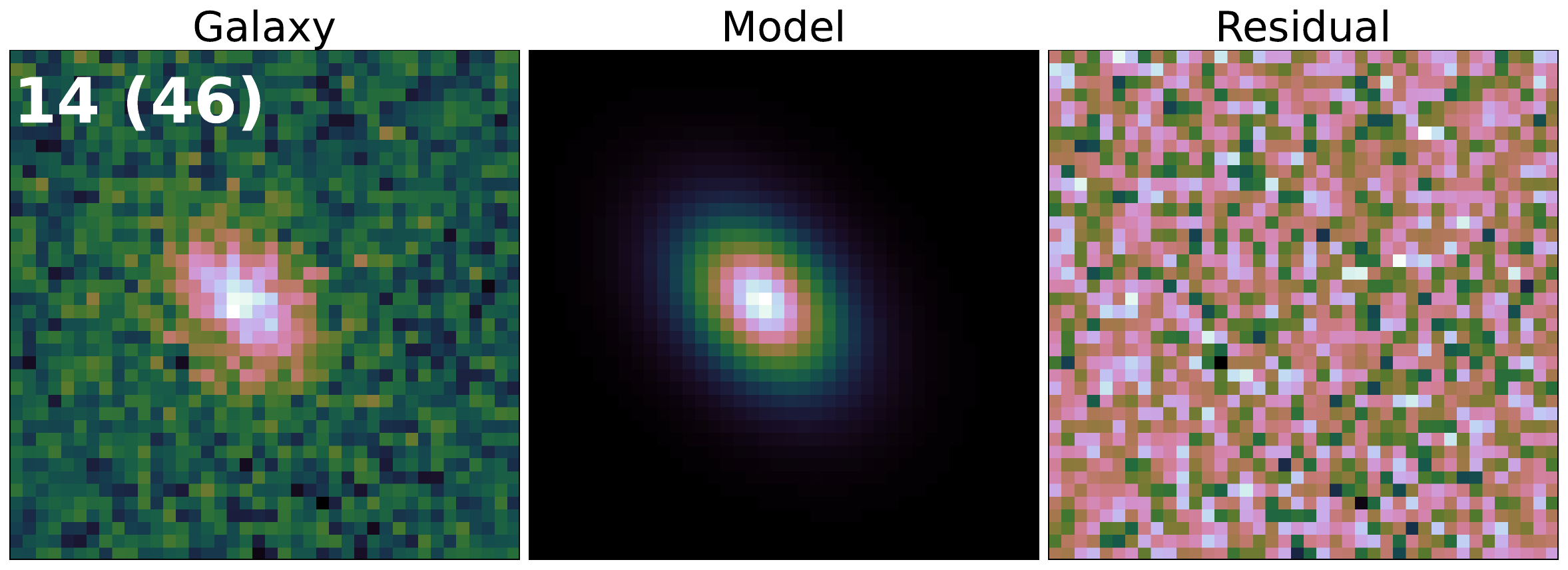}
    \end{minipage}
    
    \vspace{0.5em}
    
    \begin{minipage}{0.49\textwidth}
        \centering
        \includegraphics[width=\linewidth]{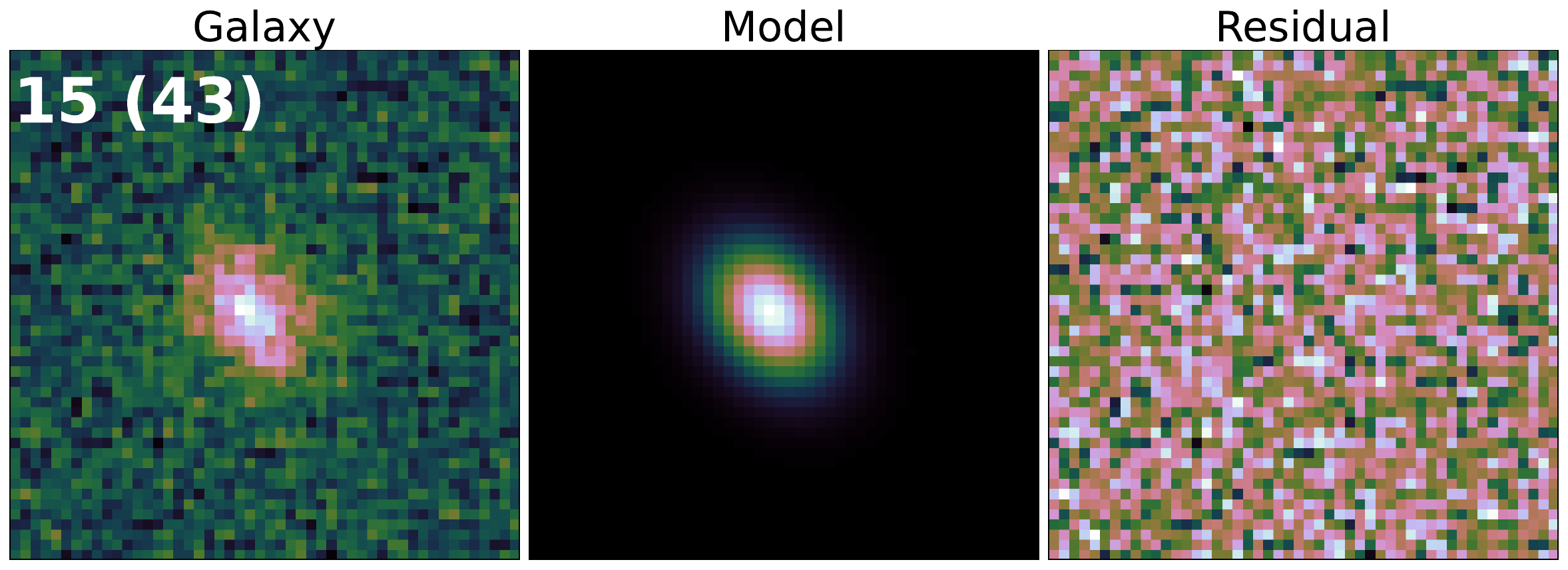}
    \end{minipage}\hfill
    \begin{minipage}{0.49\textwidth}
        \centering
        \includegraphics[width=\linewidth]{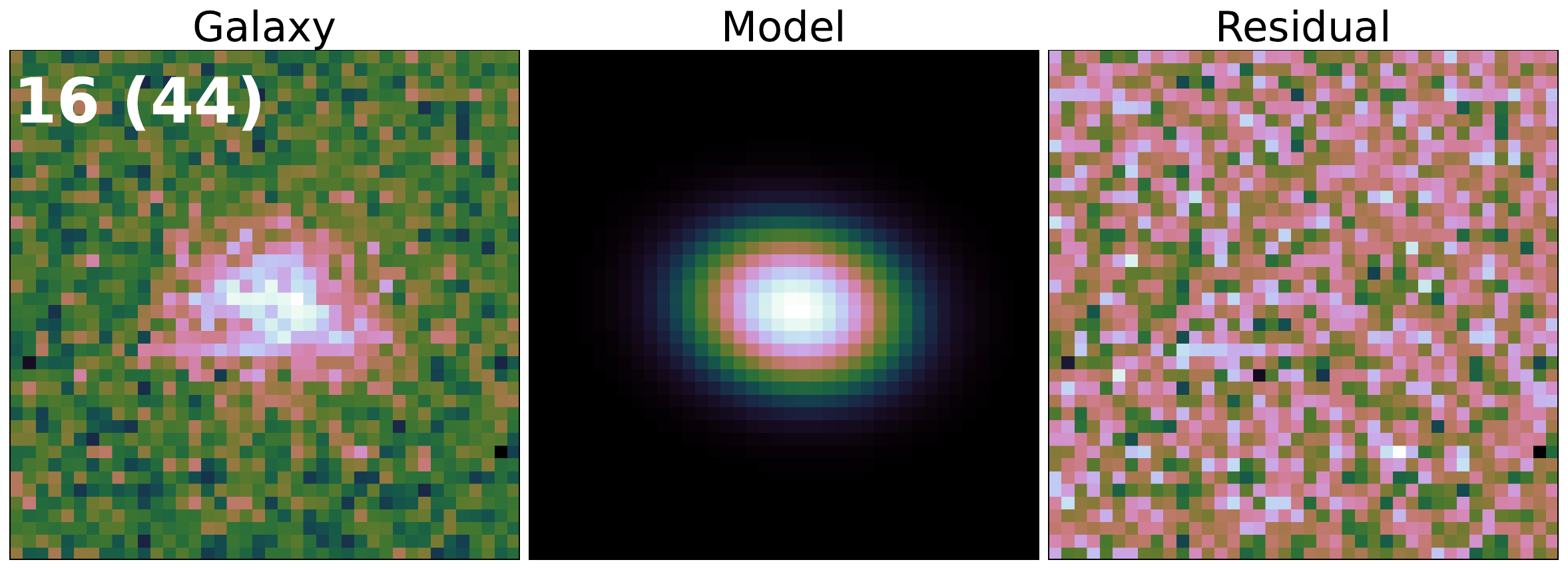}
    \end{minipage}
    
    \vspace{0.5em}

        \begin{minipage}{0.49\textwidth}
        \centering
        \includegraphics[width=\linewidth]{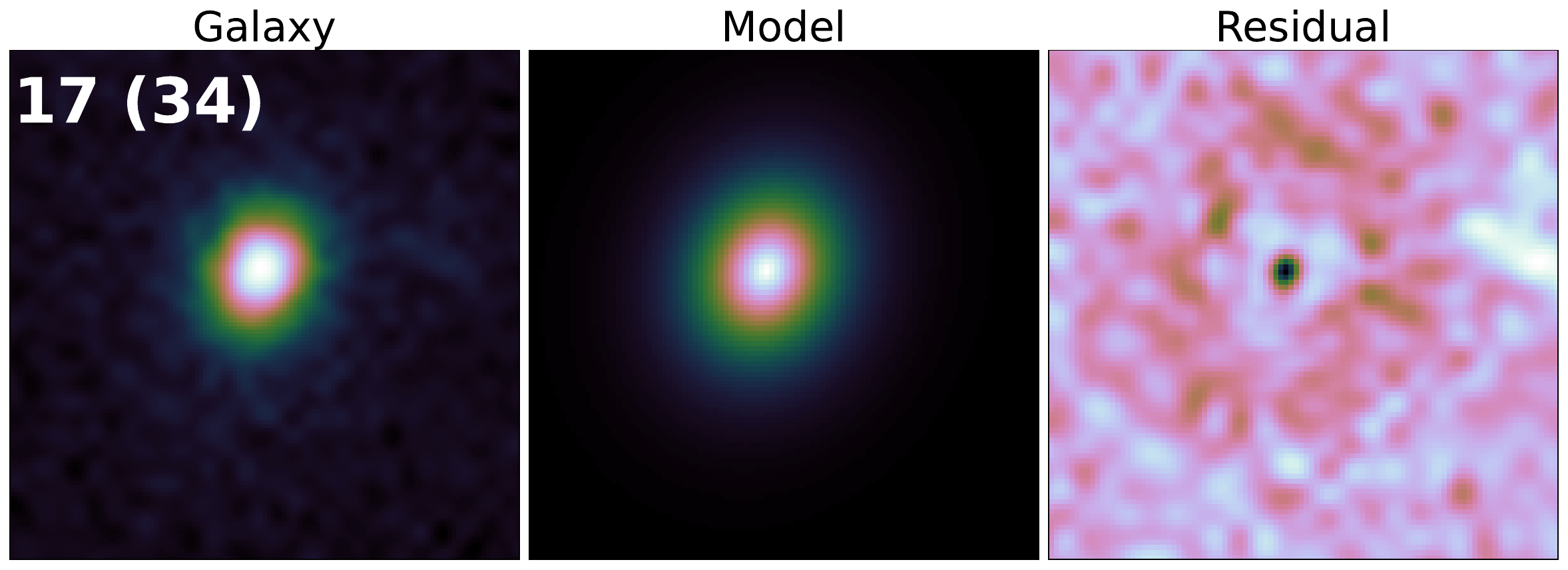}
    \end{minipage}\hfill
    \begin{minipage}{0.49\textwidth}
        \centering
        \includegraphics[width=\linewidth]{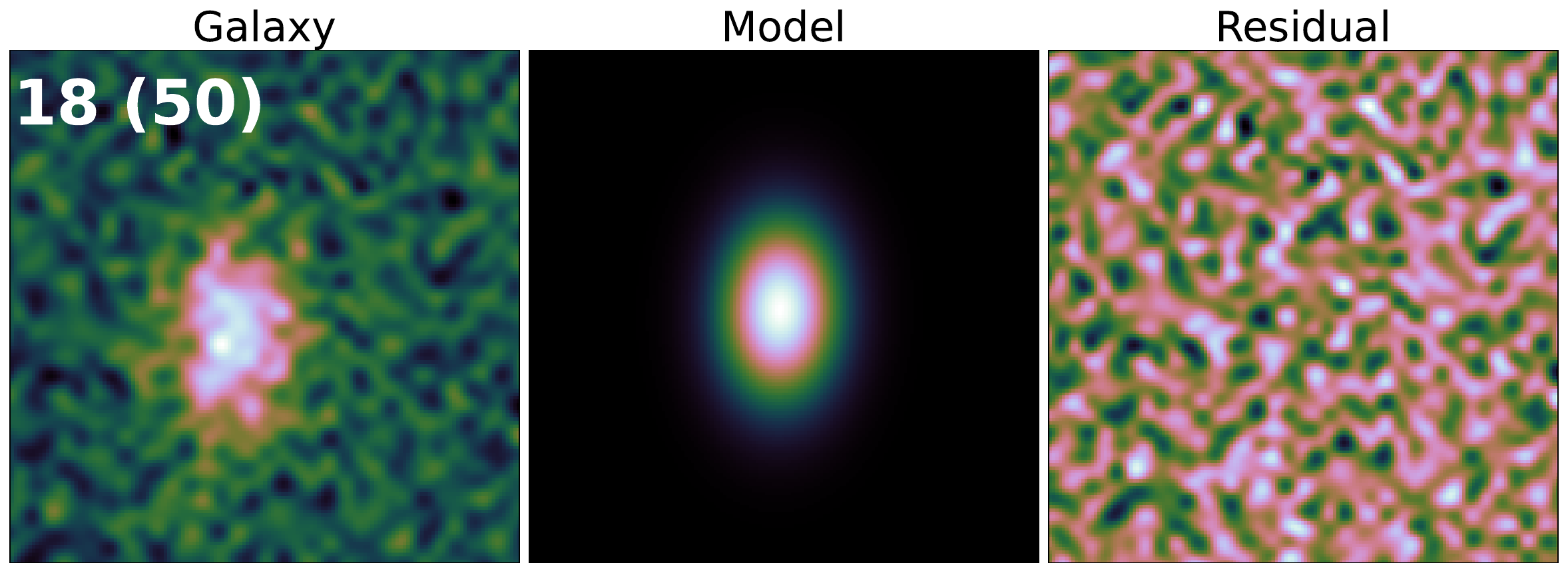}
    \end{minipage}
    
            \begin{minipage}{0.49\textwidth}
        \centering
        \includegraphics[width=\linewidth]{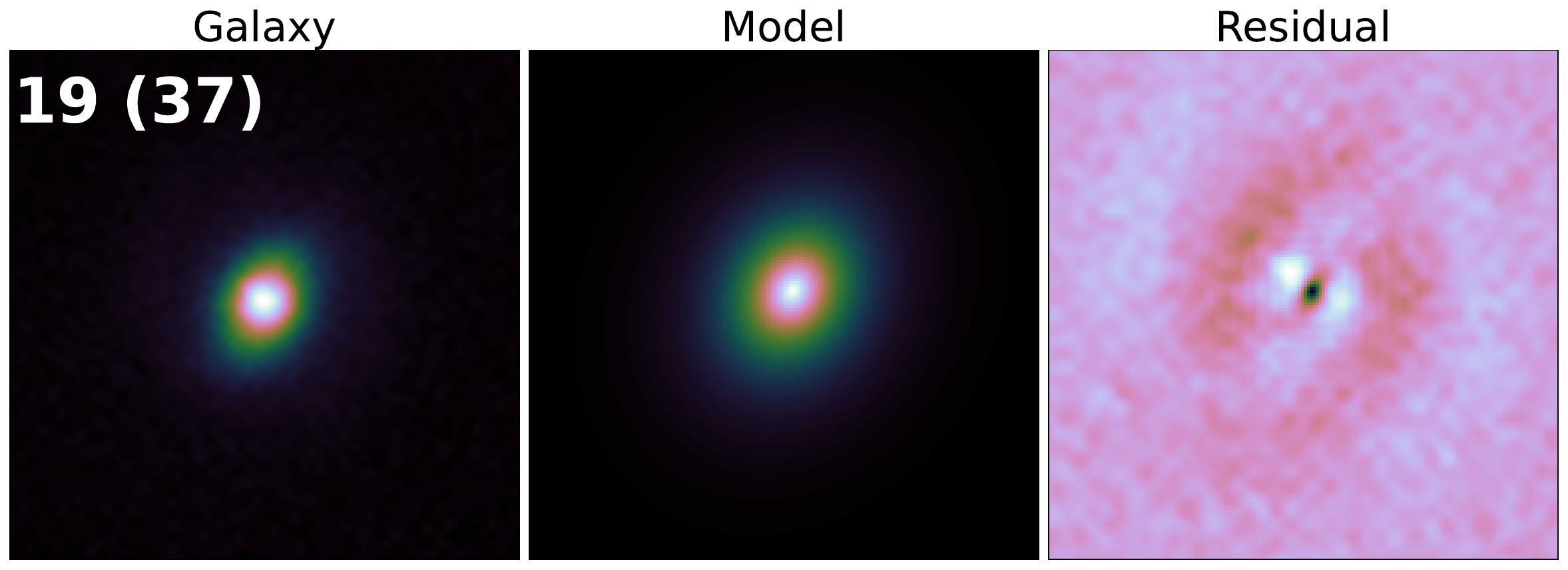}
    \end{minipage}\hfill
    \begin{minipage}{0.49\textwidth}
        \centering
        \includegraphics[width=\linewidth]{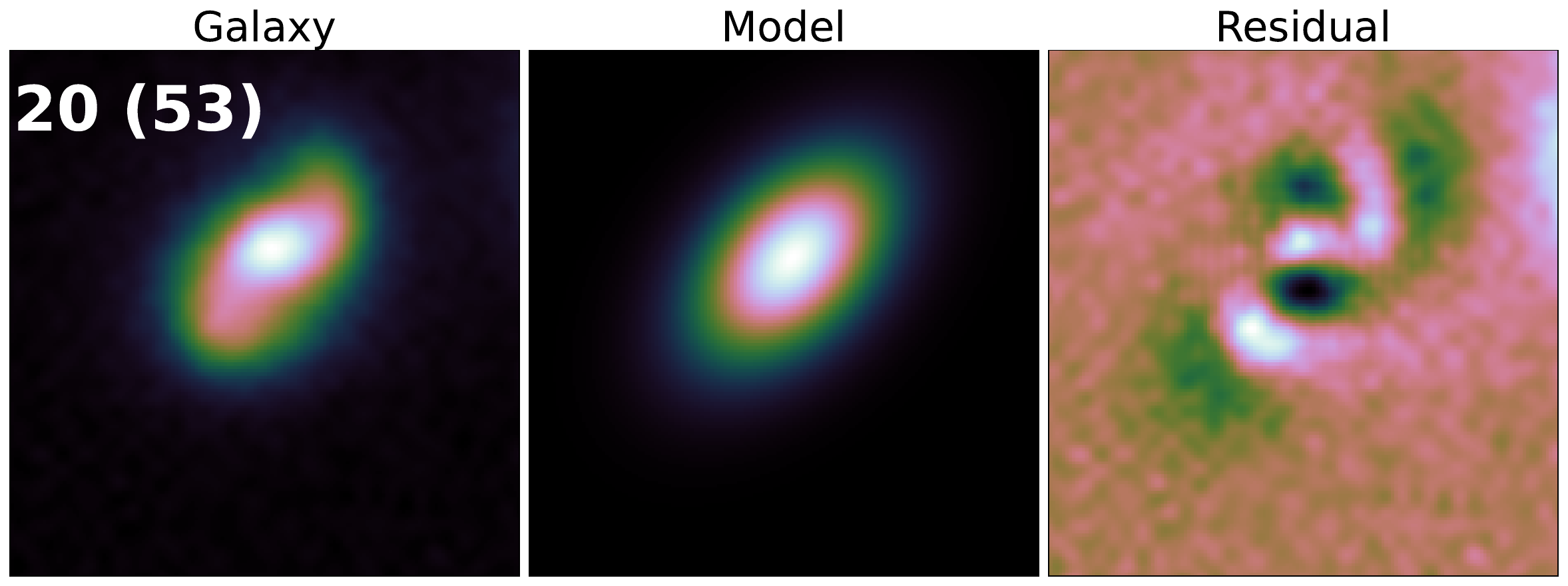}
    \end{minipage}
    
    \vspace{0.5em}
    
    \begin{minipage}{0.49\textwidth}
        \centering
        \includegraphics[width=\linewidth]{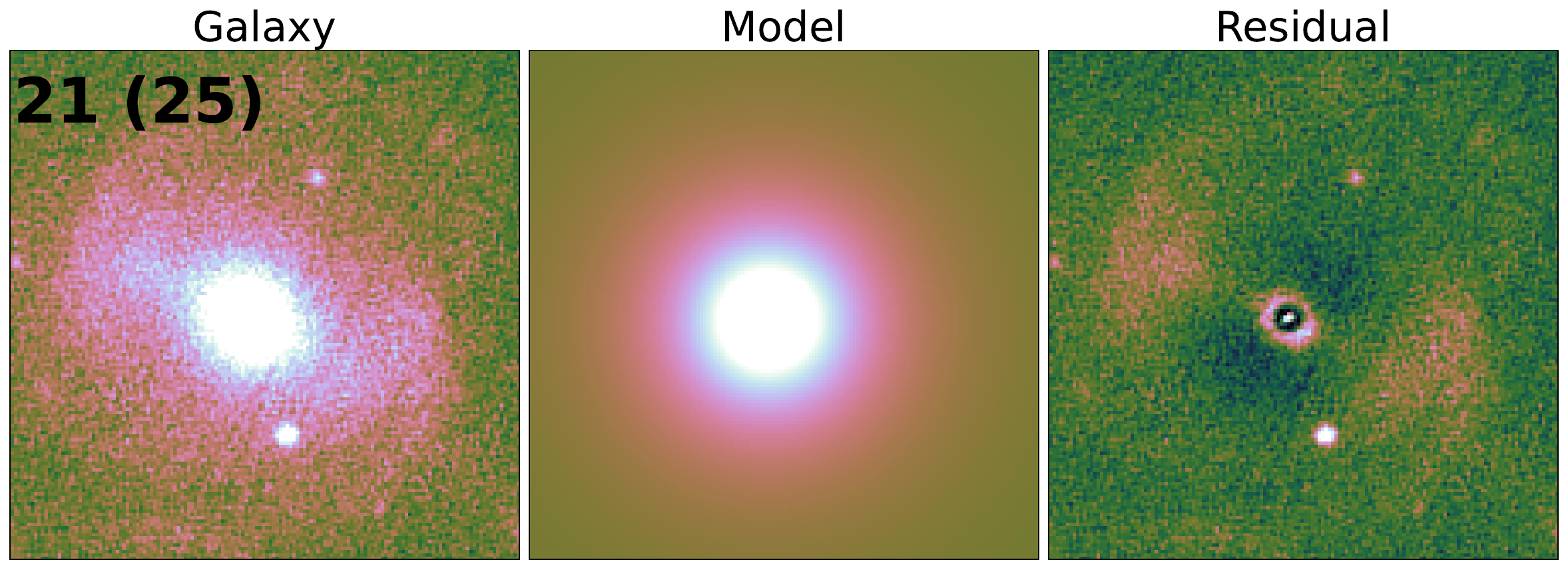}
    \end{minipage}\hfill
    \begin{minipage}{0.49\textwidth}
        \centering
        \includegraphics[width=\linewidth]{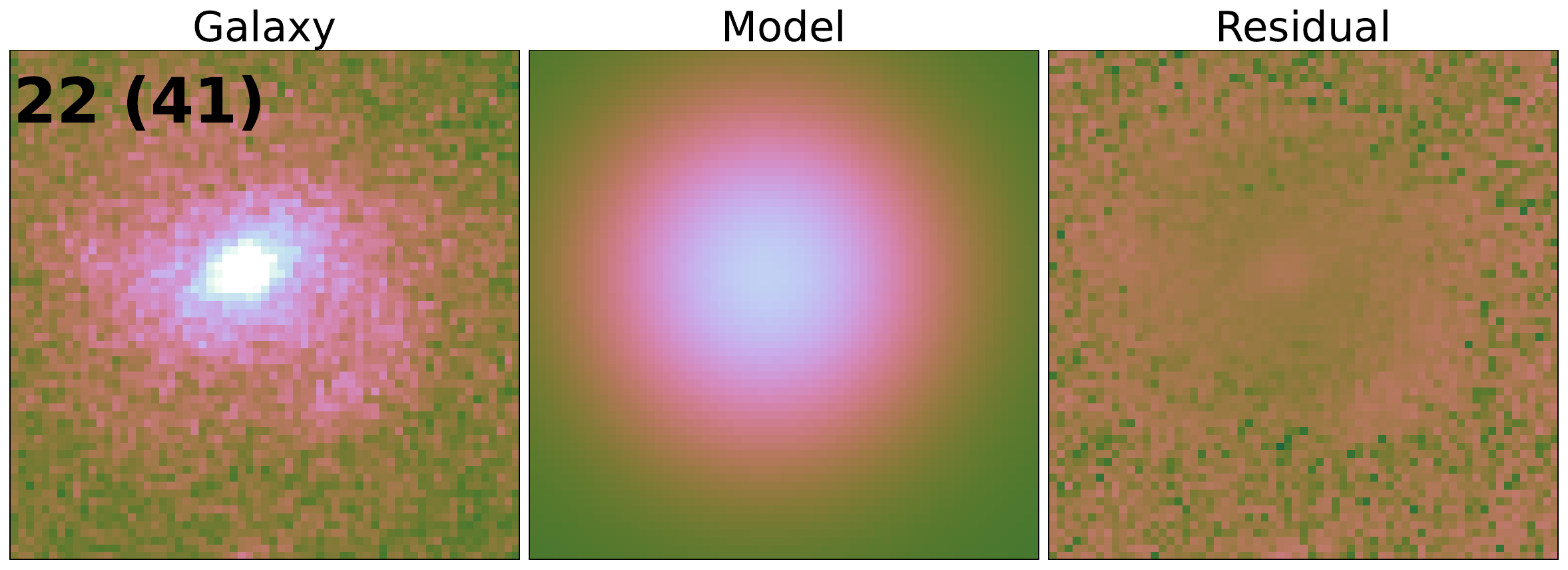}
    \end{minipage}
    
    \vspace{0.5em}
    
        \begin{minipage}{0.49\textwidth}
        \centering
        \includegraphics[width=\linewidth]{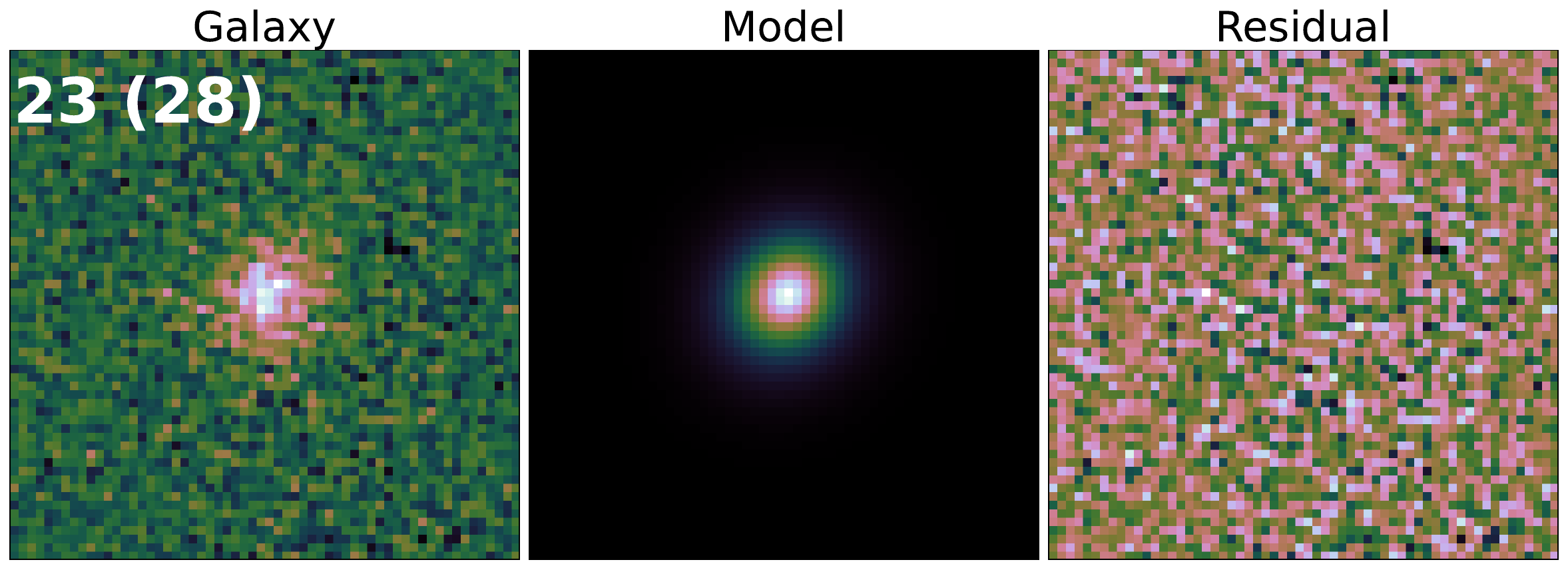}
    \end{minipage}\hfill

    \vspace{0.5em}
\end{figure*}

\pagebreak
\begin{figure*}
    \caption{Non-parametric profile fitting results for 23 FRB host galaxies using the \texttt{Photutils} package: For each galaxy, we show three images - (left) r-band image of the host, (center) best-fitted galaxy model, and (right) resultant residual. The numbers in the top left of each r-band image correspond to the host number and the estimated inclination angle  (in degrees, quoted in parentheses), as shown in Table \ref{tab:host_universe_frbs}. The X-axis and Y-axis of all plots represent right ascension and declination, respectively.}\label{fig:photutils}
    \centering
    
    \begin{minipage}{0.49\textwidth}
        \centering
        \includegraphics[width=\linewidth]{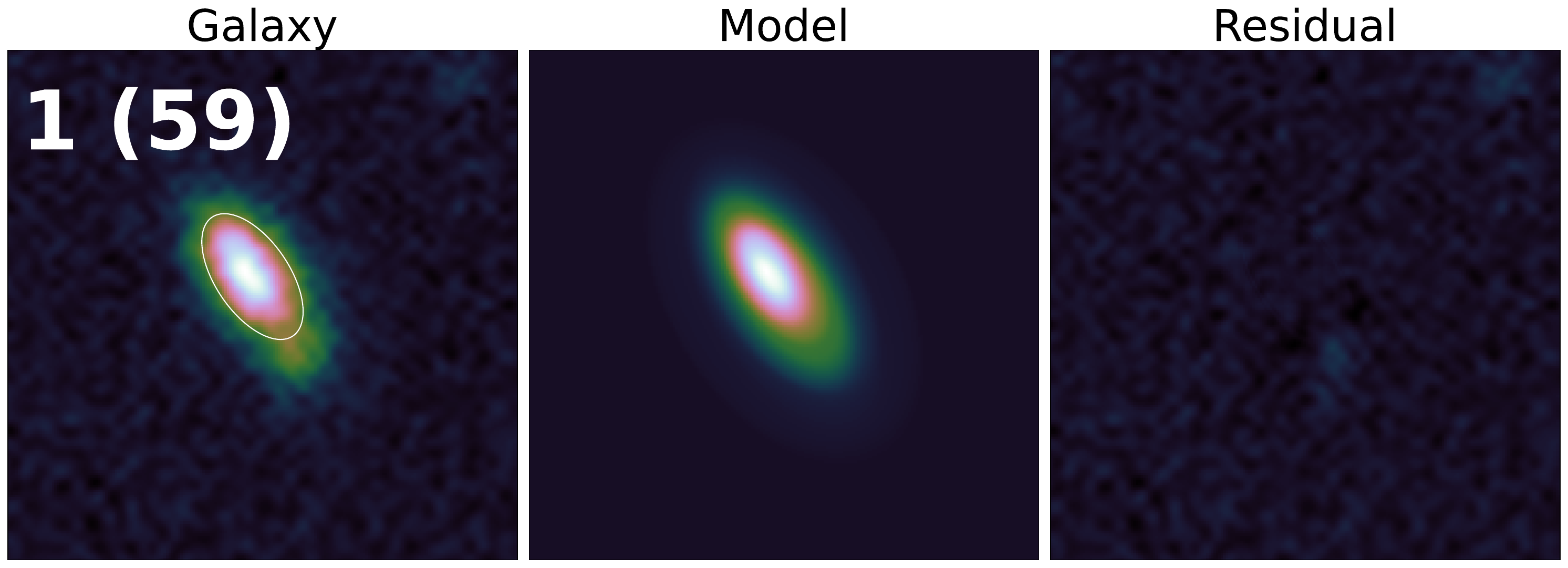}
    \end{minipage}\hfill
    \begin{minipage}{0.49\textwidth}
        \centering
        \includegraphics[width=\linewidth]{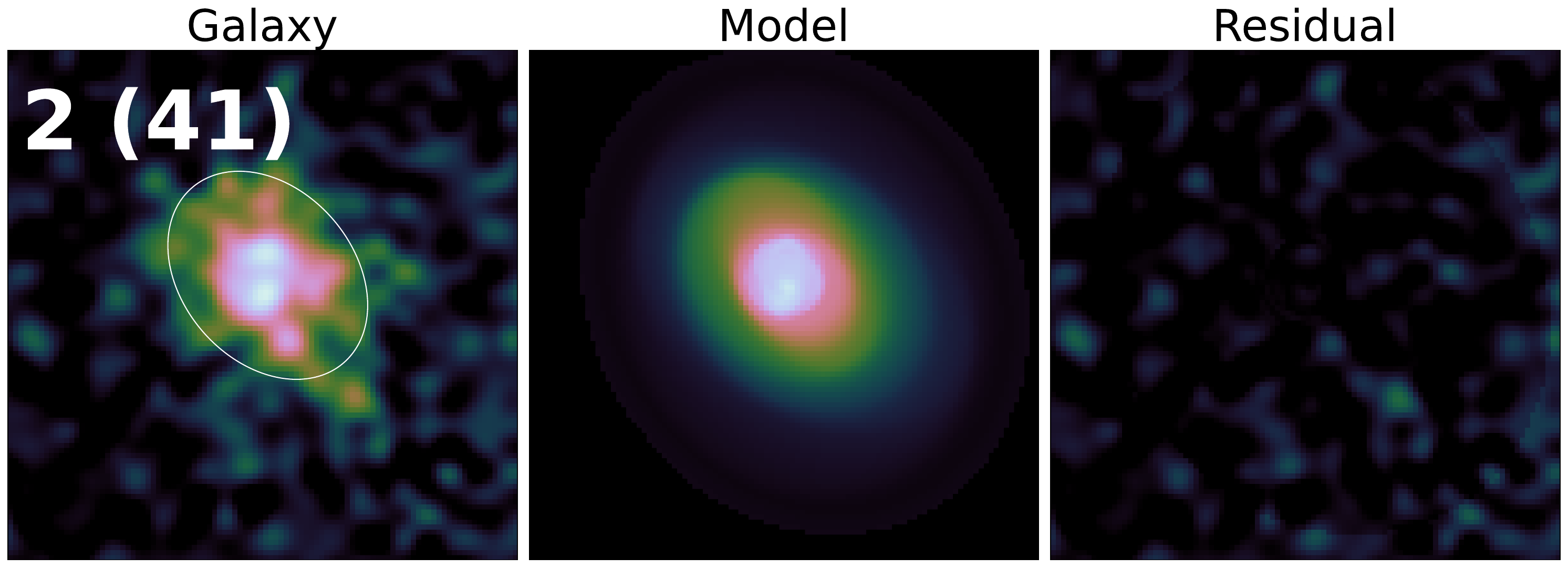}
    \end{minipage}
    
    \vspace{0.5em}
    
    \begin{minipage}{0.49\textwidth}
        \centering
        \includegraphics[width=\linewidth]{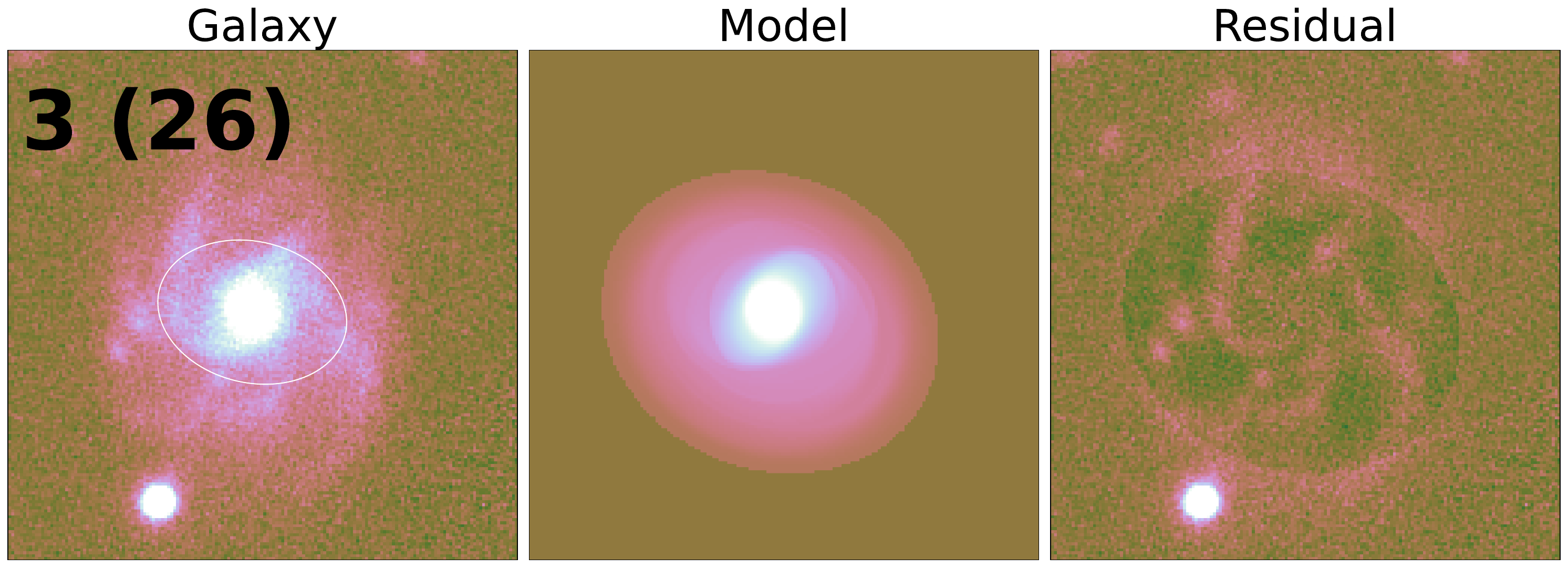}
    \end{minipage}\hfill
    \begin{minipage}{0.49\textwidth}
        \centering
        \includegraphics[width=\linewidth]{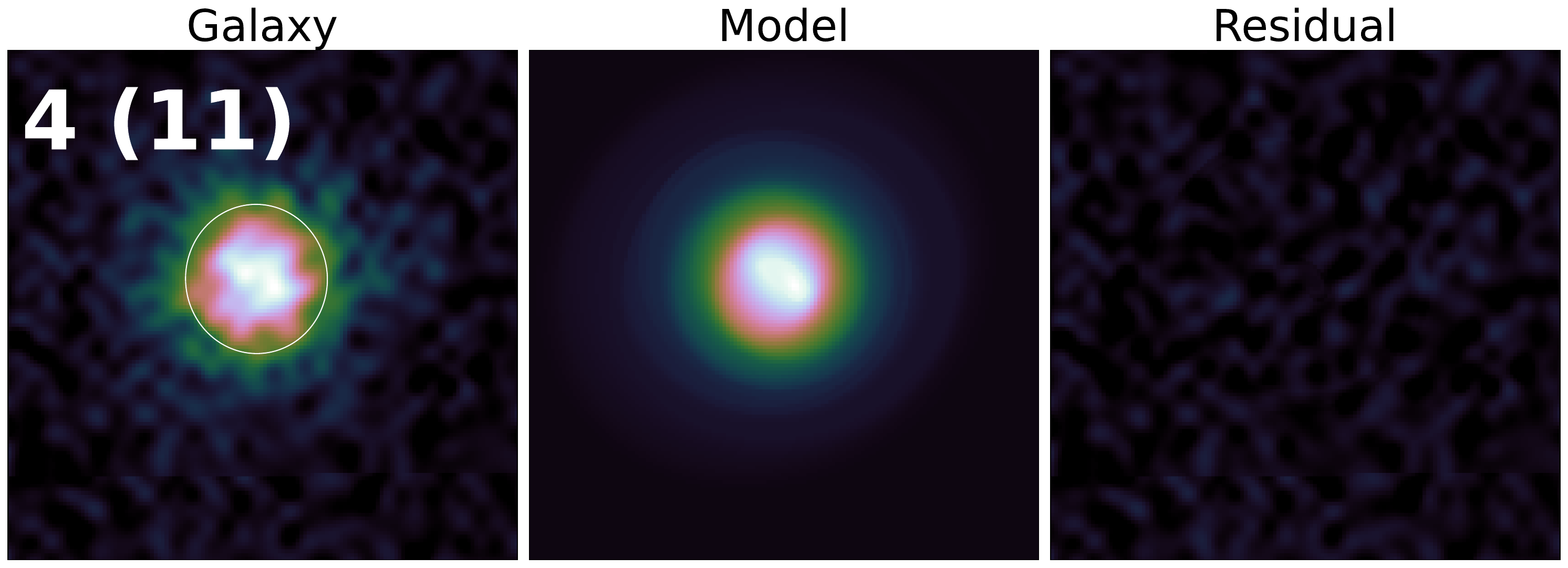}
    \end{minipage}
    
    \vspace{0.5em}
    
        \begin{minipage}{0.49\textwidth}
        \centering
        \includegraphics[width=\linewidth]{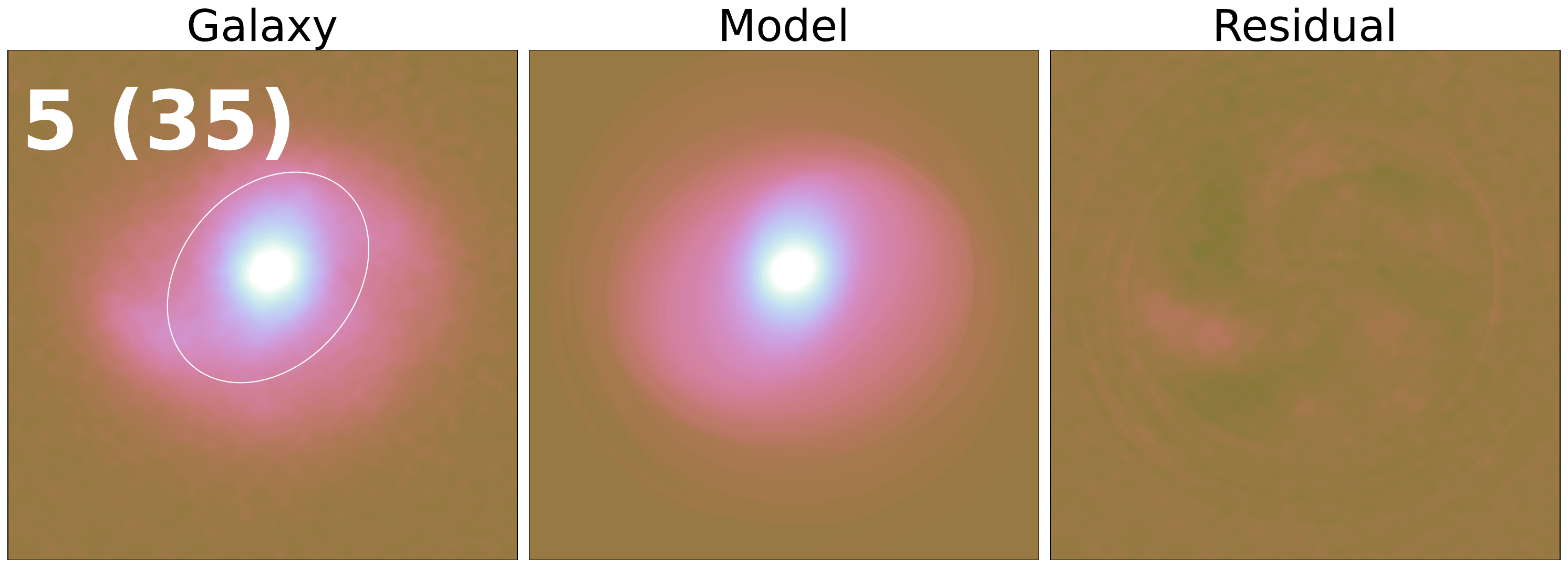}
    \end{minipage}\hfill
    \begin{minipage}{0.49\textwidth}
        \centering
        \includegraphics[width=\linewidth]{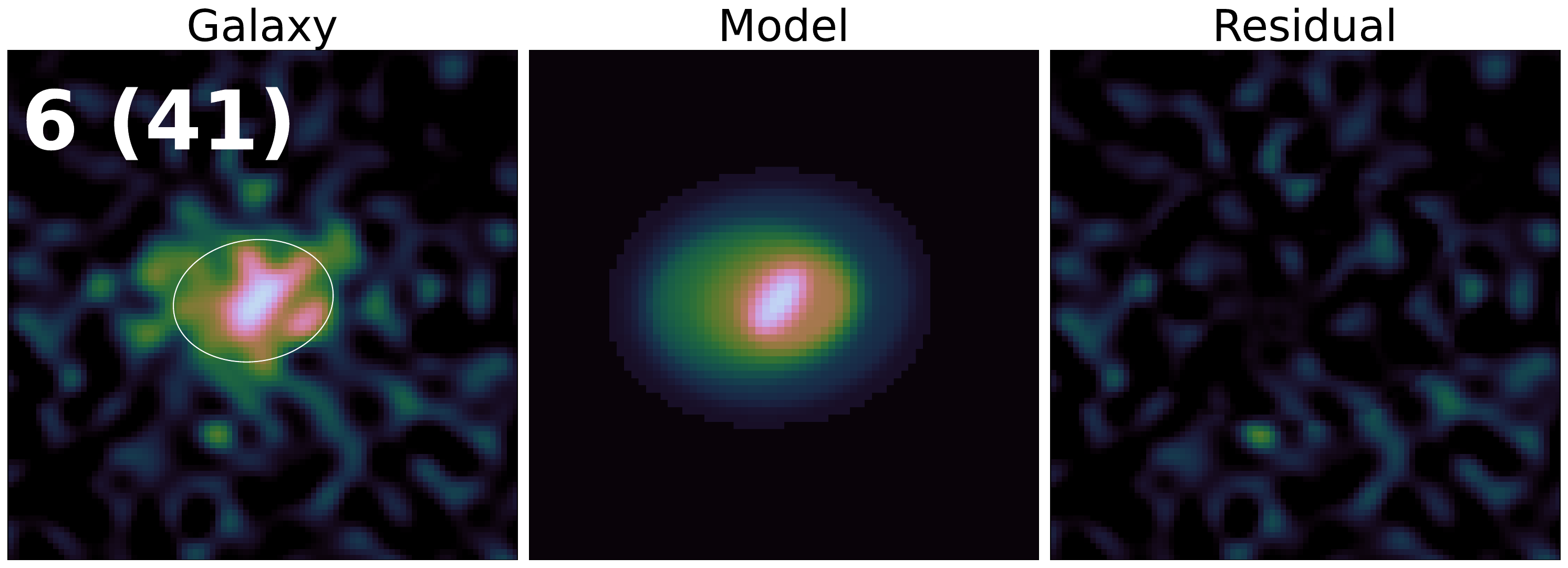}
    \end{minipage}
    
    \vspace{0.5em}
    
        \begin{minipage}{0.49\textwidth}
        \centering
        \includegraphics[width=\linewidth]{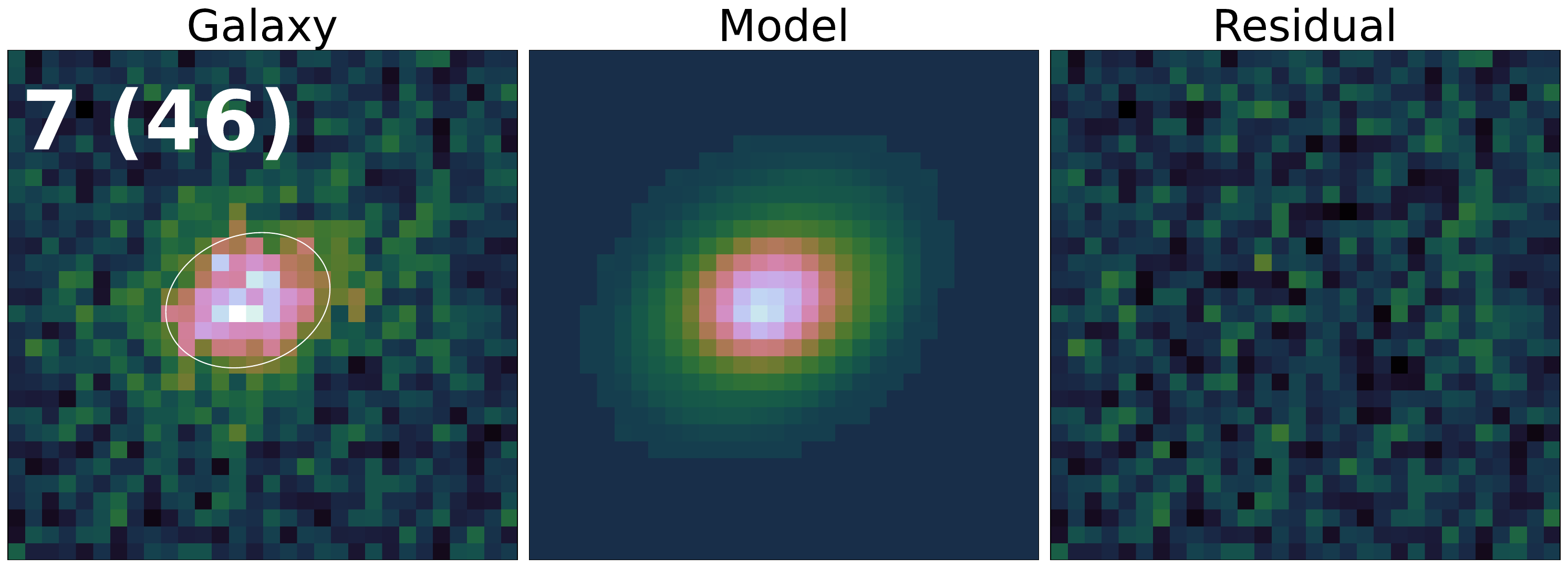}
    \end{minipage}\hfill
    \begin{minipage}{0.49\textwidth}
        \centering
        \includegraphics[width=\linewidth]{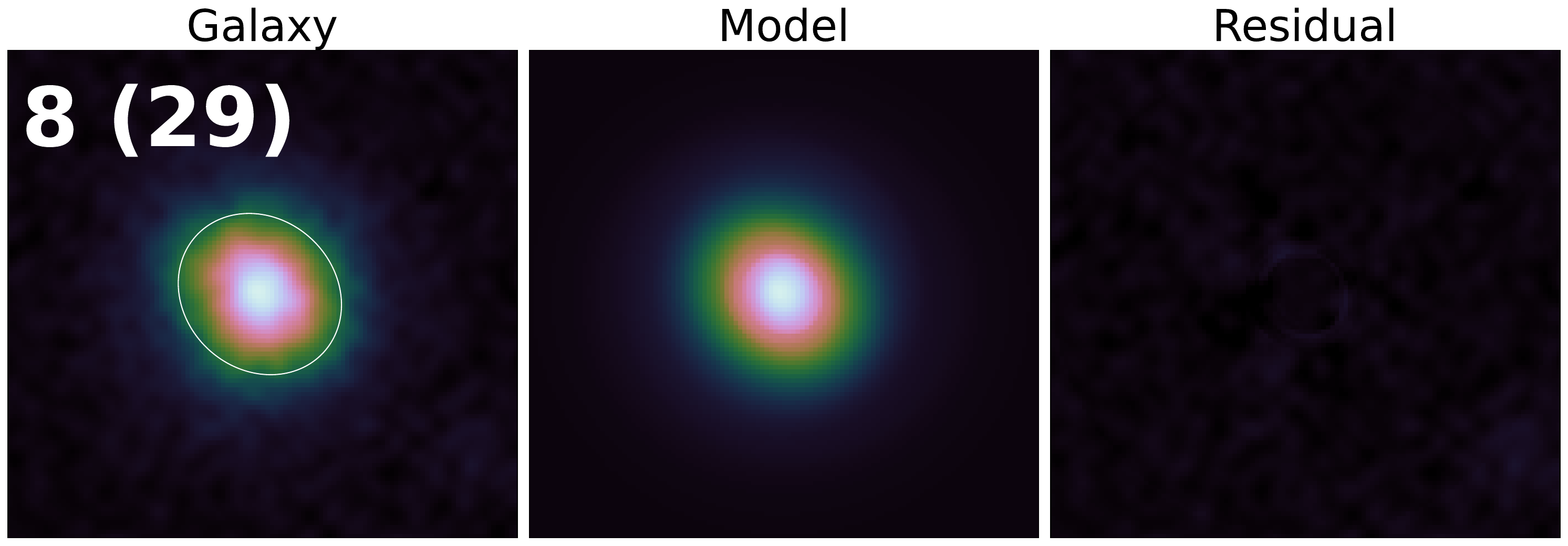}
    \end{minipage}
    
    \vspace{0.5em}
    
    \begin{minipage}{0.49\textwidth}
        \centering
        \includegraphics[width=\linewidth]{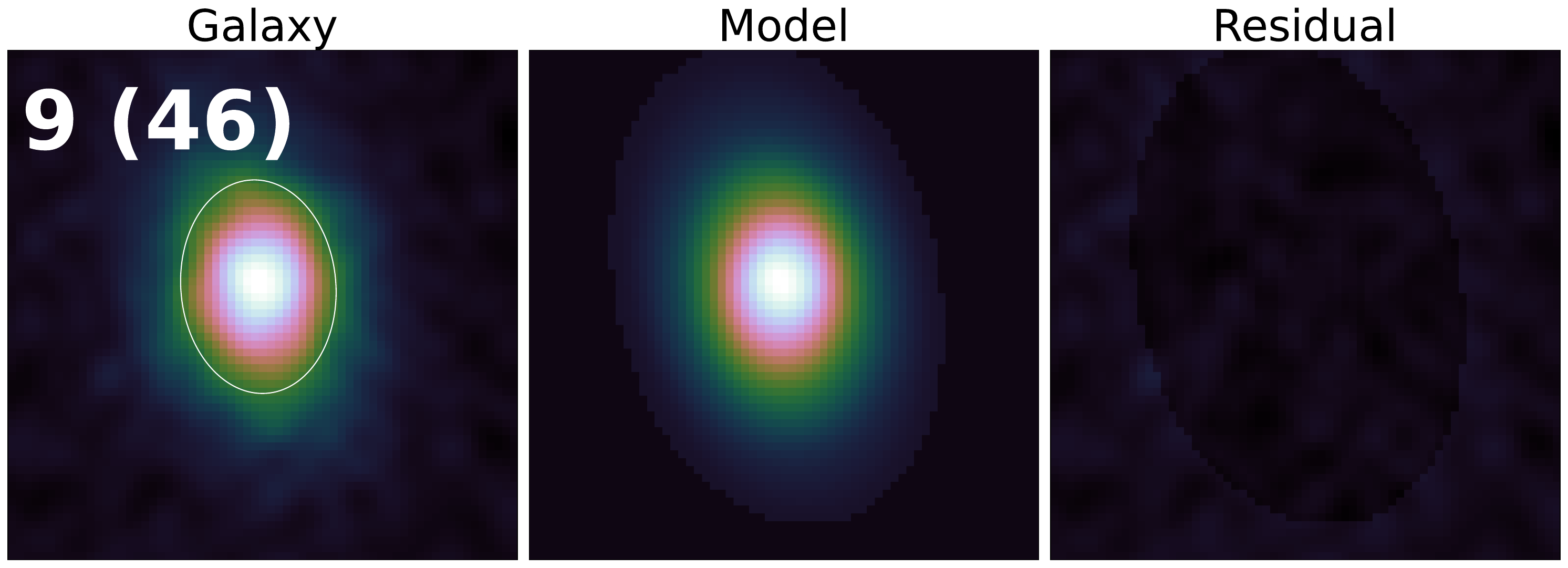}
    \end{minipage}\hfill
    \begin{minipage}{0.49\textwidth}
        \centering
        \includegraphics[width=\linewidth]{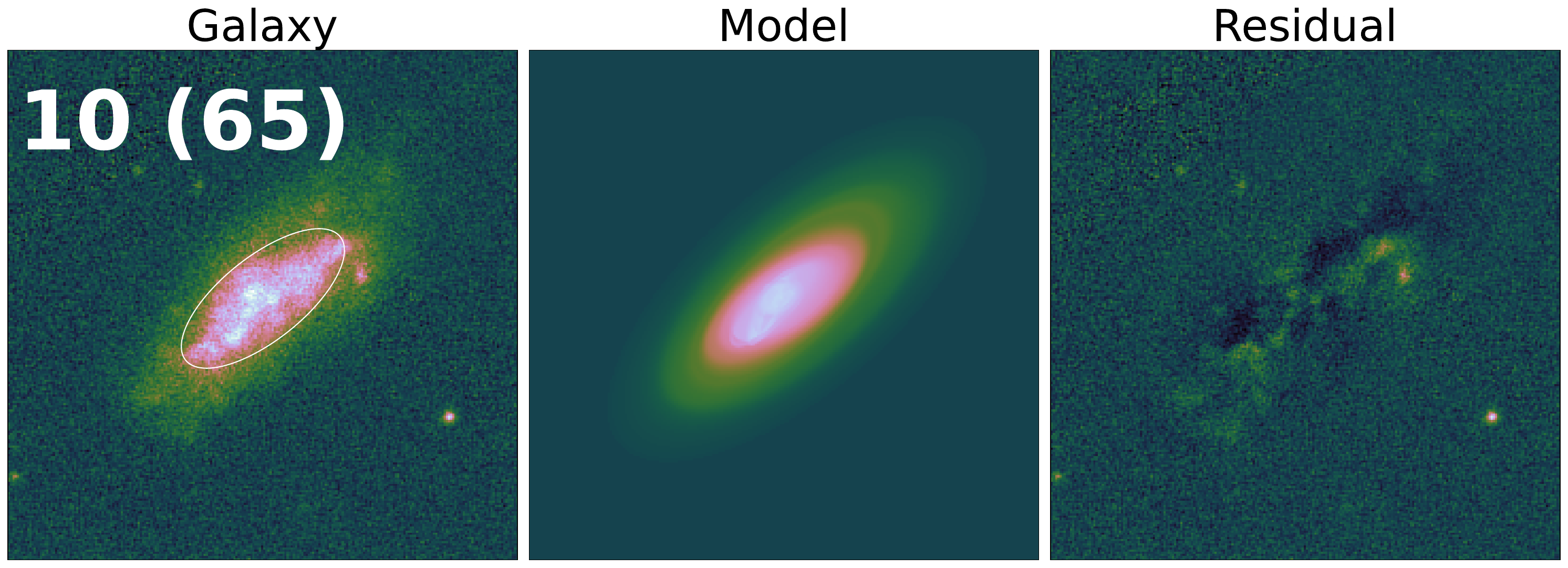}
    \end{minipage}
    
    \vspace{0.5em}
    
        \begin{minipage}{0.49\textwidth}
        \centering
        \includegraphics[width=\linewidth]{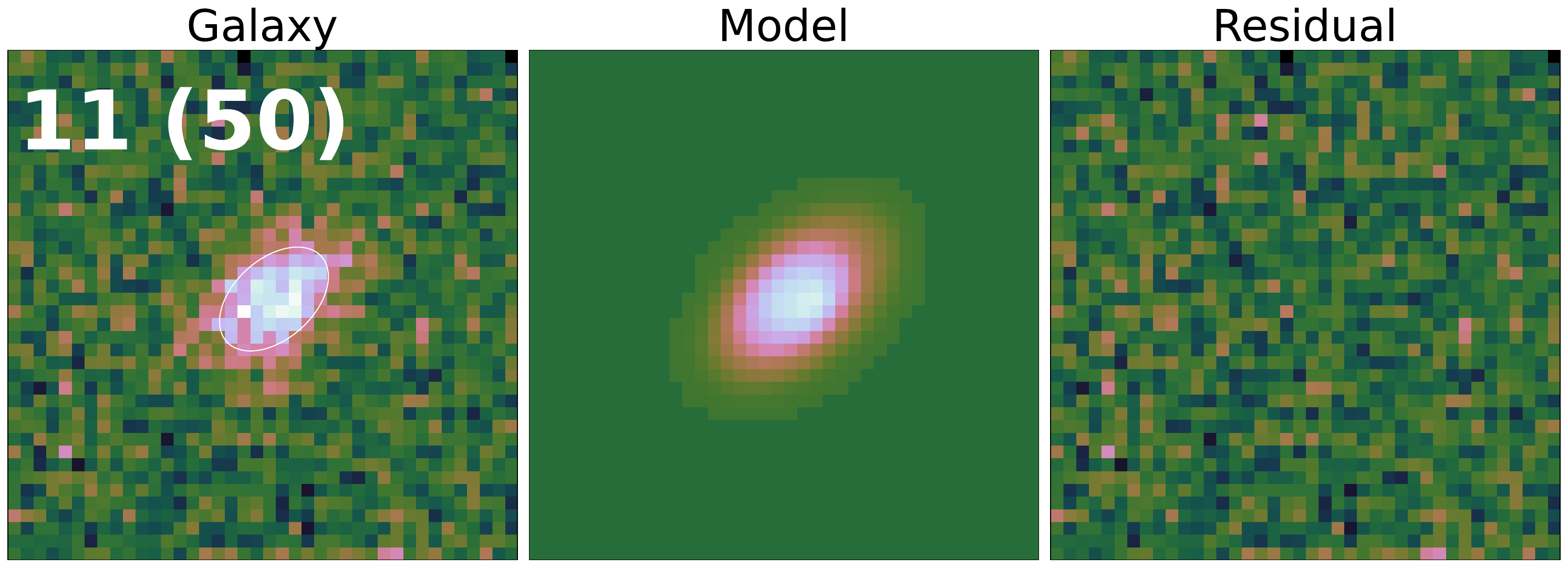}
    \end{minipage}\hfill
    \begin{minipage}{0.49\textwidth}
        \centering
        \includegraphics[width=\linewidth]{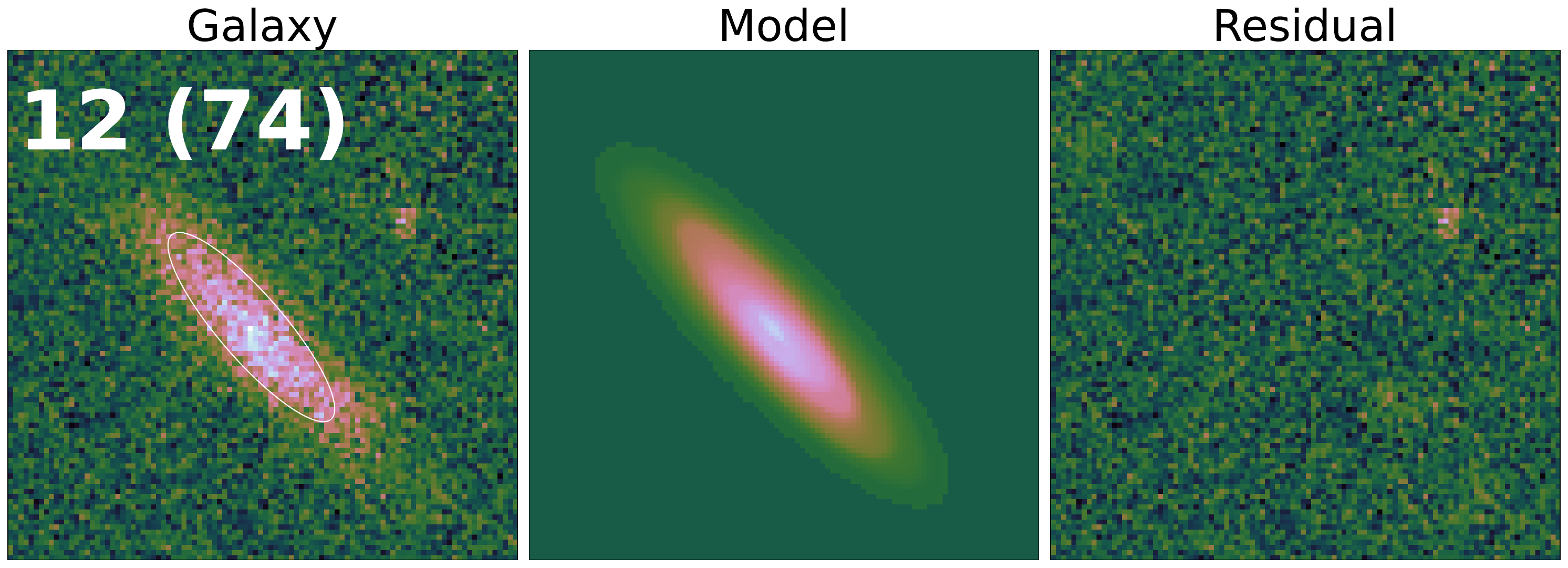}
    \end{minipage}

    \phantomcaption
\end{figure*}

\begin{figure*}
\ContinuedFloat
    \vspace{0.5em}
    
            \begin{minipage}{0.49\textwidth}
        \centering
        \includegraphics[width=\linewidth]{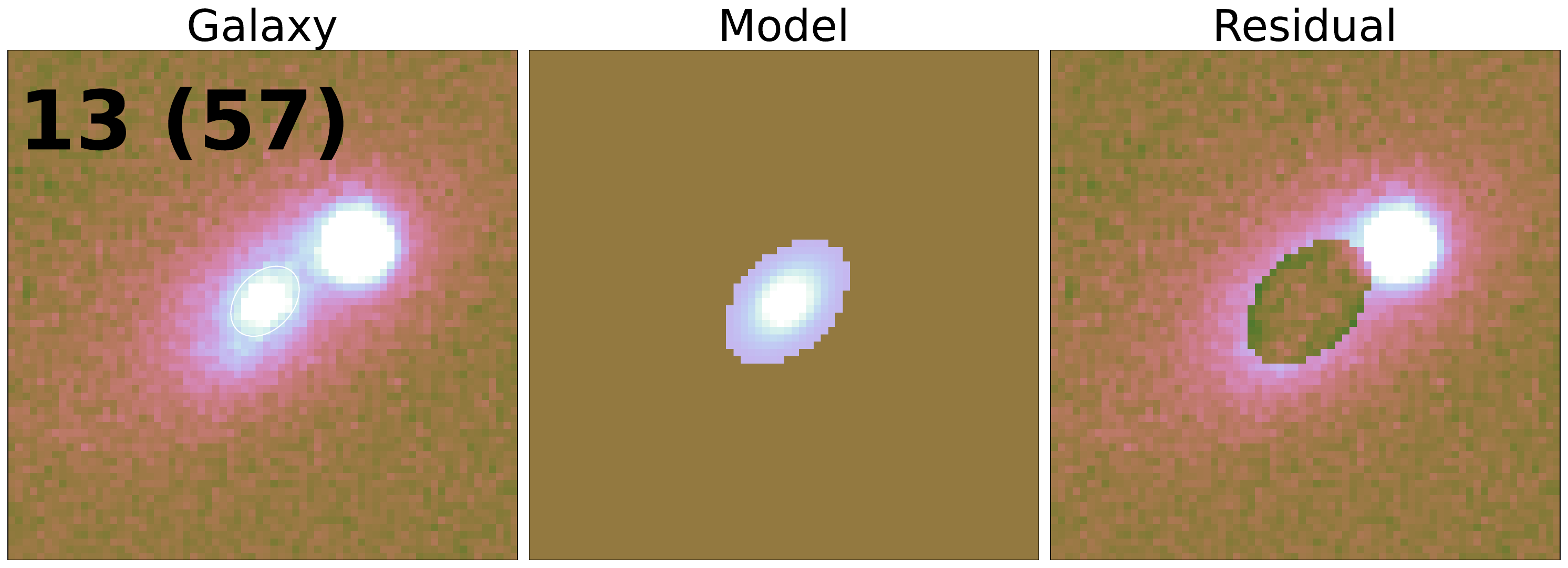}
    \end{minipage}\hfill
    \begin{minipage}{0.49\textwidth}
        \centering
        \includegraphics[width=\linewidth]{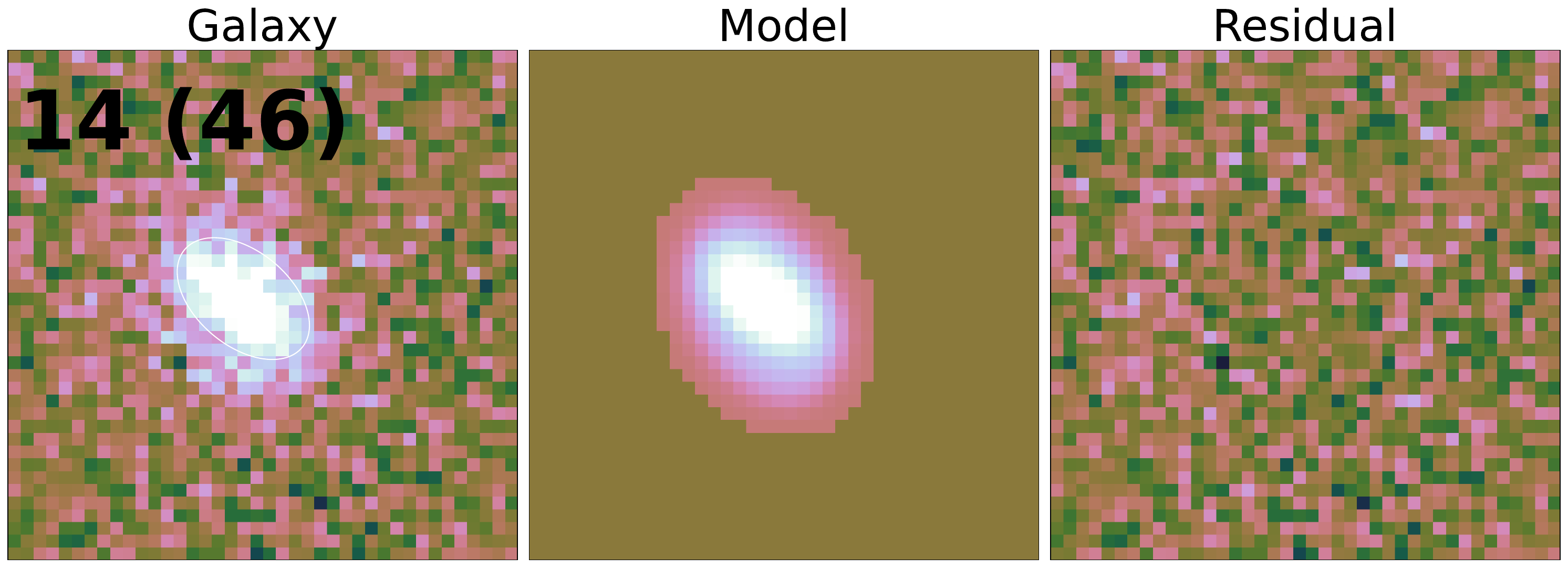}
    \end{minipage}
    
    \vspace{0.5em}
    
    \begin{minipage}{0.49\textwidth}
        \centering
        \includegraphics[width=\linewidth]{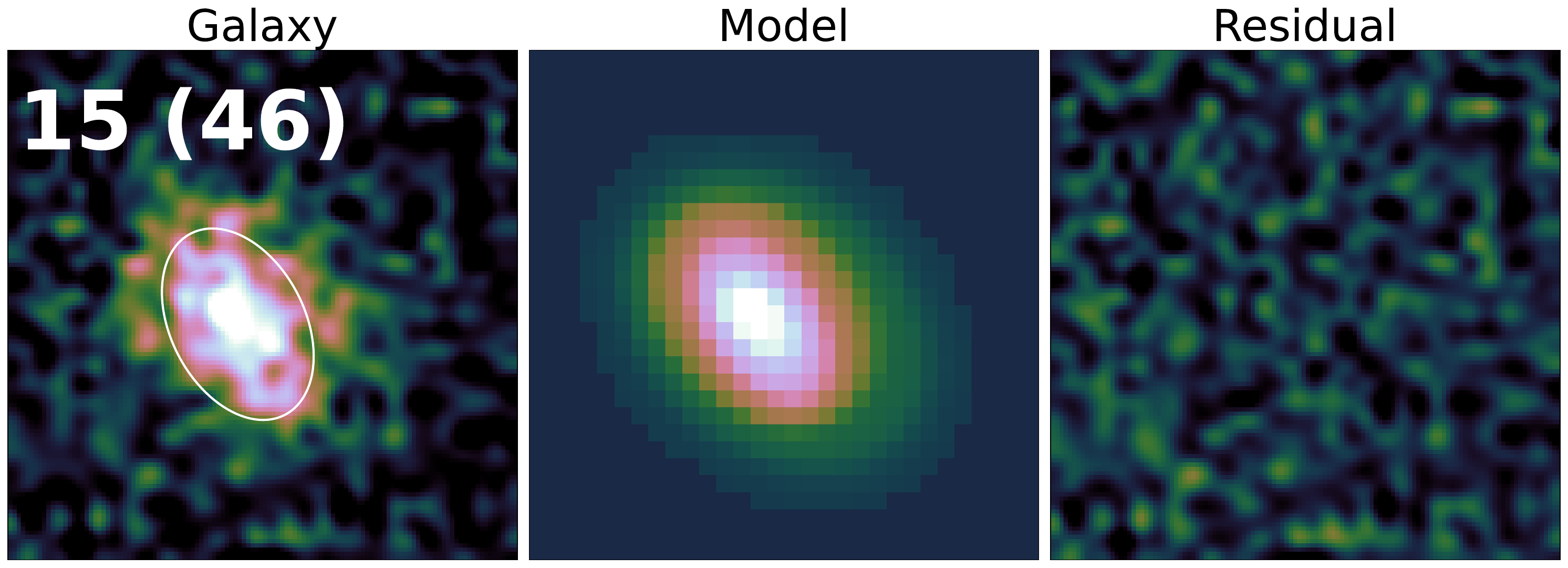}
    \end{minipage}\hfill
    \begin{minipage}{0.49\textwidth}
        \centering
        \includegraphics[width=\linewidth]{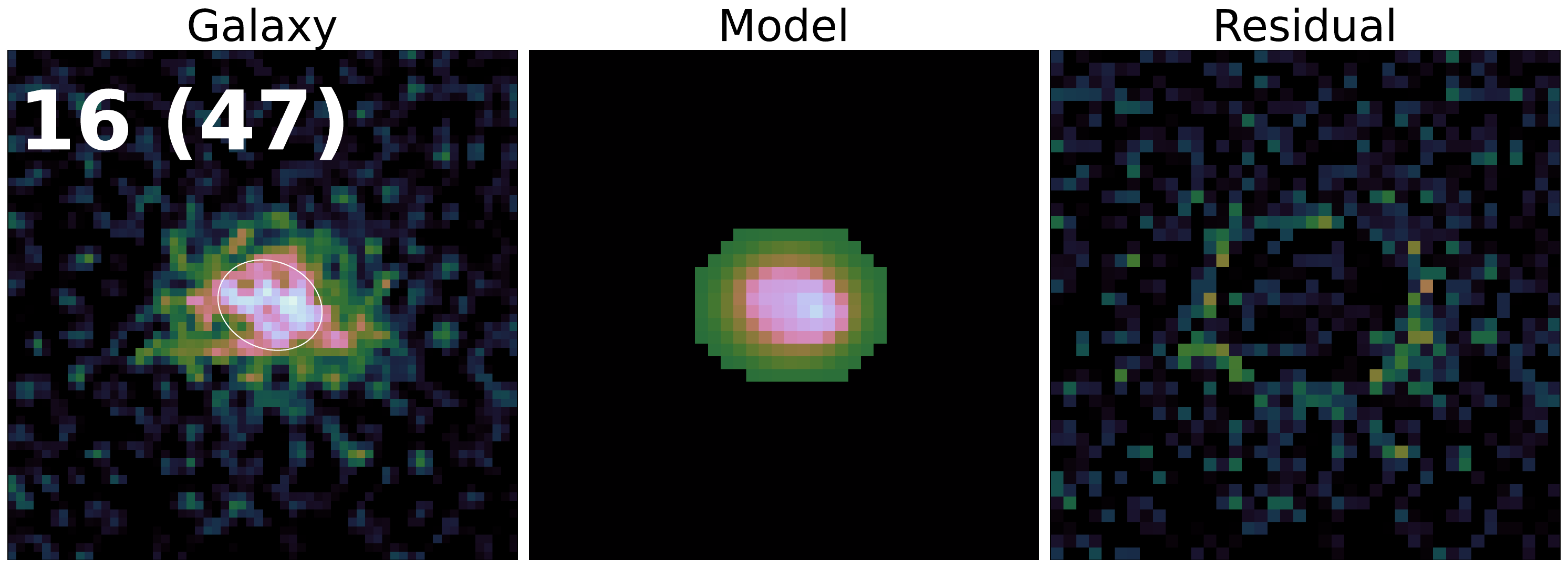}
    \end{minipage}
    
    \vspace{0.5em}

        \begin{minipage}{0.49\textwidth}
        \centering
        \includegraphics[width=\linewidth]{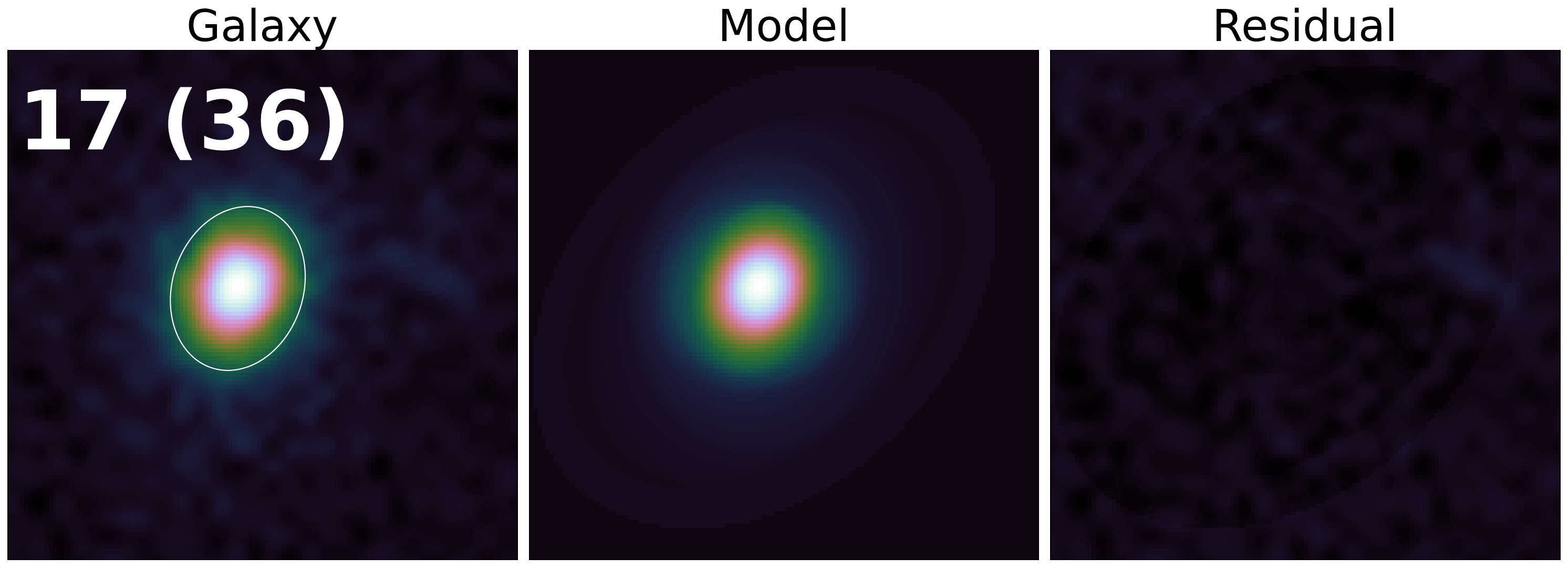}
    \end{minipage}\hfill
    \begin{minipage}{0.49\textwidth}
        \centering
        \includegraphics[width=\linewidth]{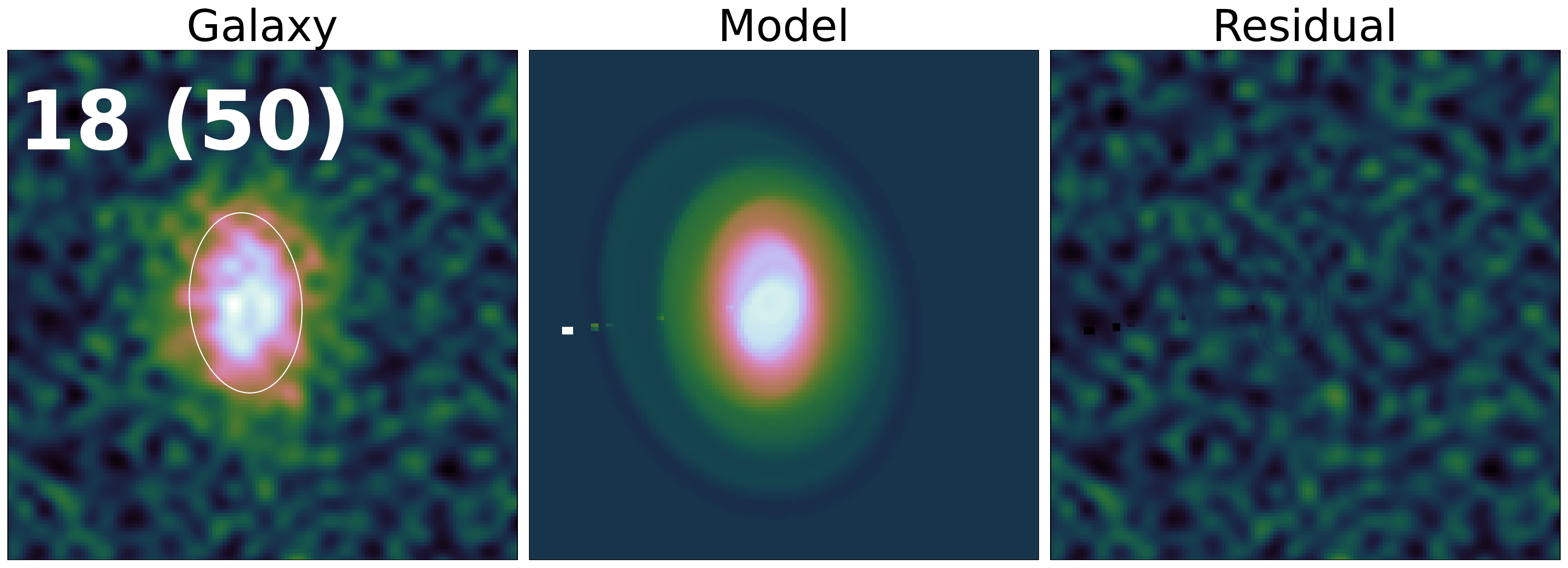}
    \end{minipage}
    
            \begin{minipage}{0.49\textwidth}
        \centering
        \includegraphics[width=\linewidth]{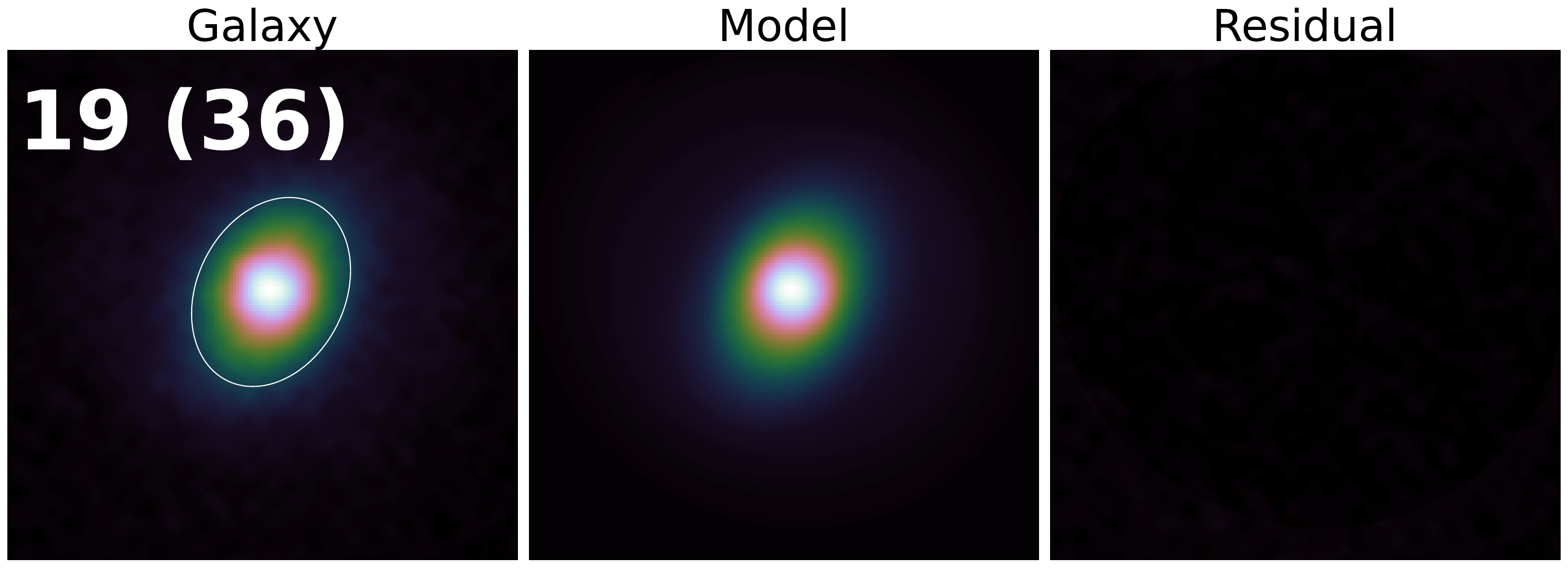}
    \end{minipage}\hfill
    \begin{minipage}{0.49\textwidth}
        \centering
        \includegraphics[width=\linewidth]{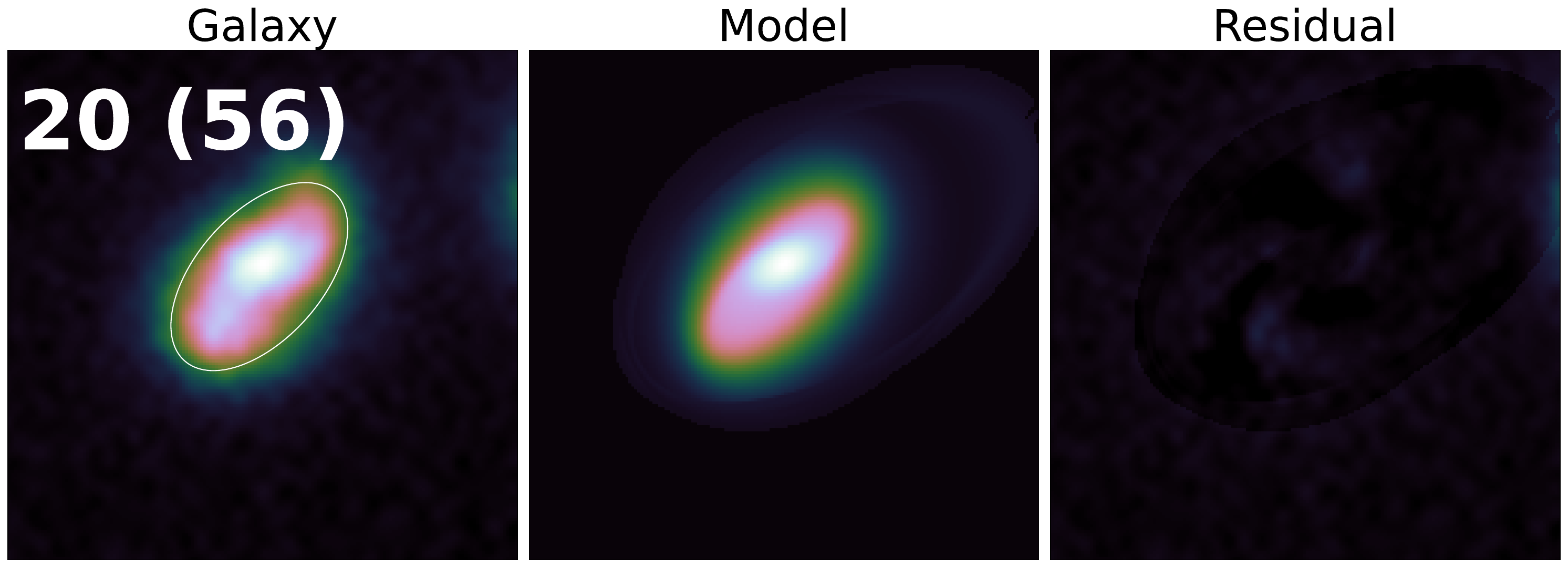}
    \end{minipage}
    
    \vspace{0.5em}
    
    \begin{minipage}{0.49\textwidth}
        \centering
        \includegraphics[width=\linewidth]{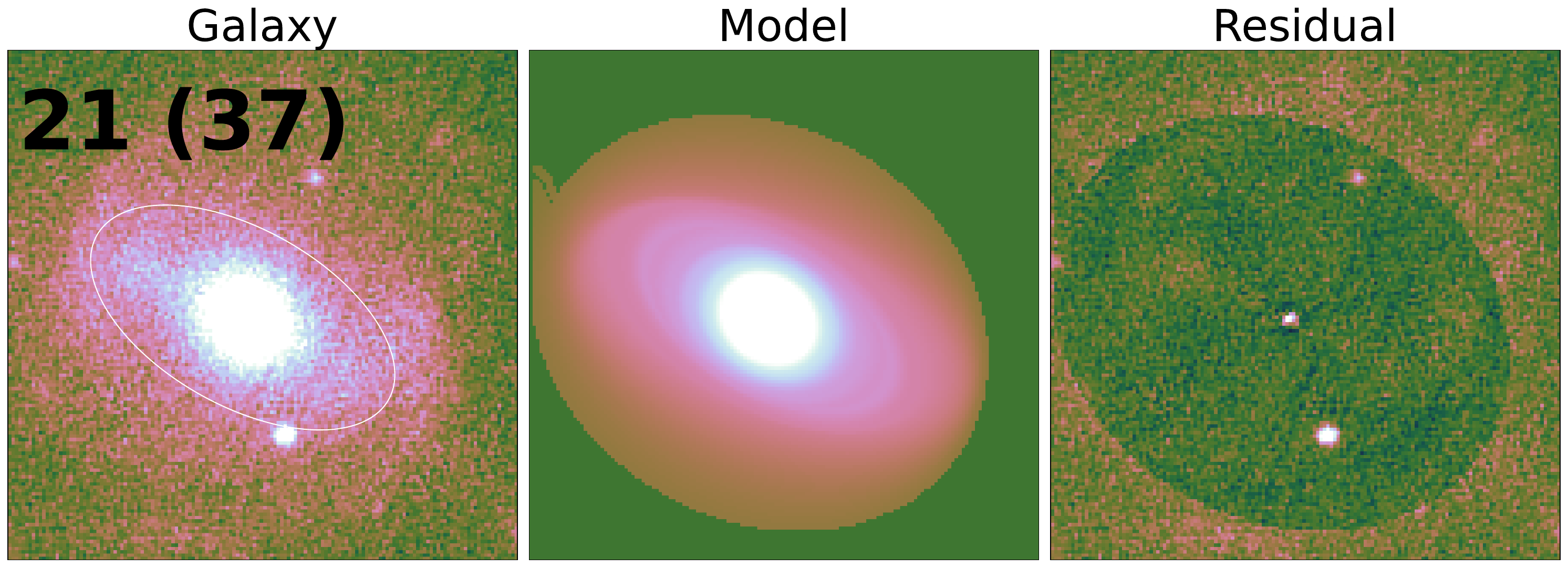}
    \end{minipage}\hfill
    \begin{minipage}{0.49\textwidth}
        \centering
        \includegraphics[width=\linewidth]{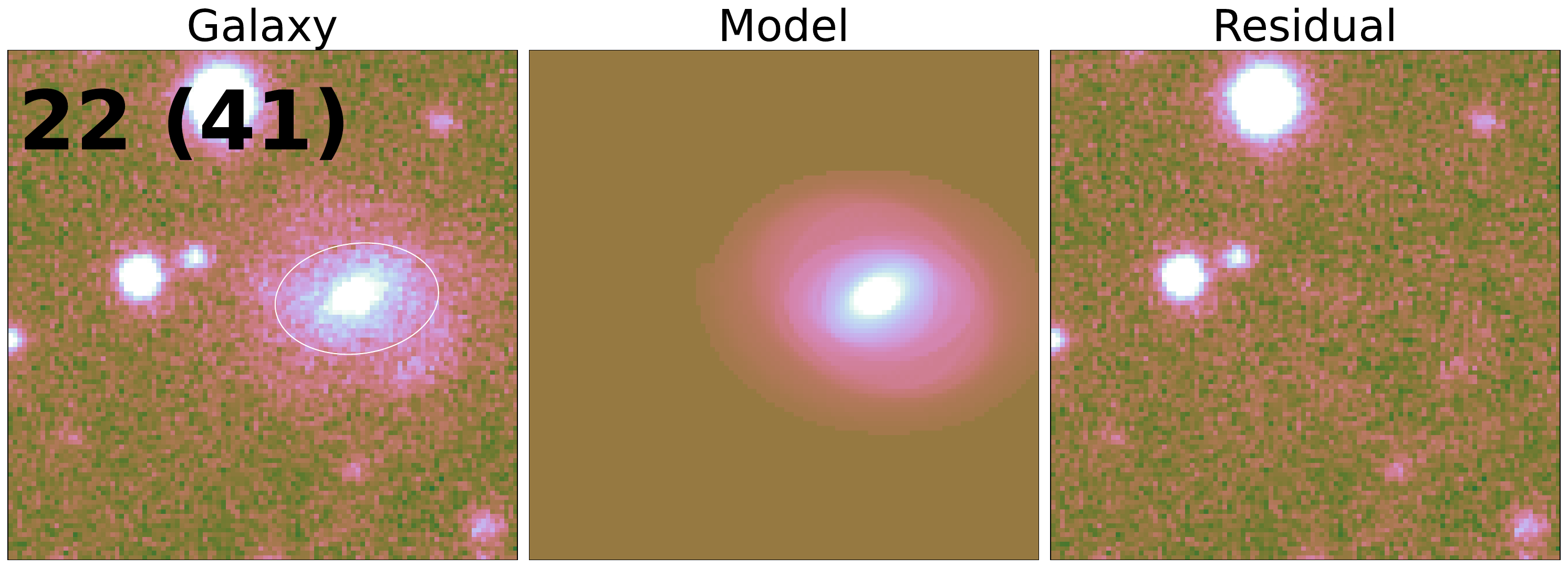}
    \end{minipage}
    
    \vspace{0.5em}
    
        \begin{minipage}{0.49\textwidth}
        \centering
        \includegraphics[width=\linewidth]{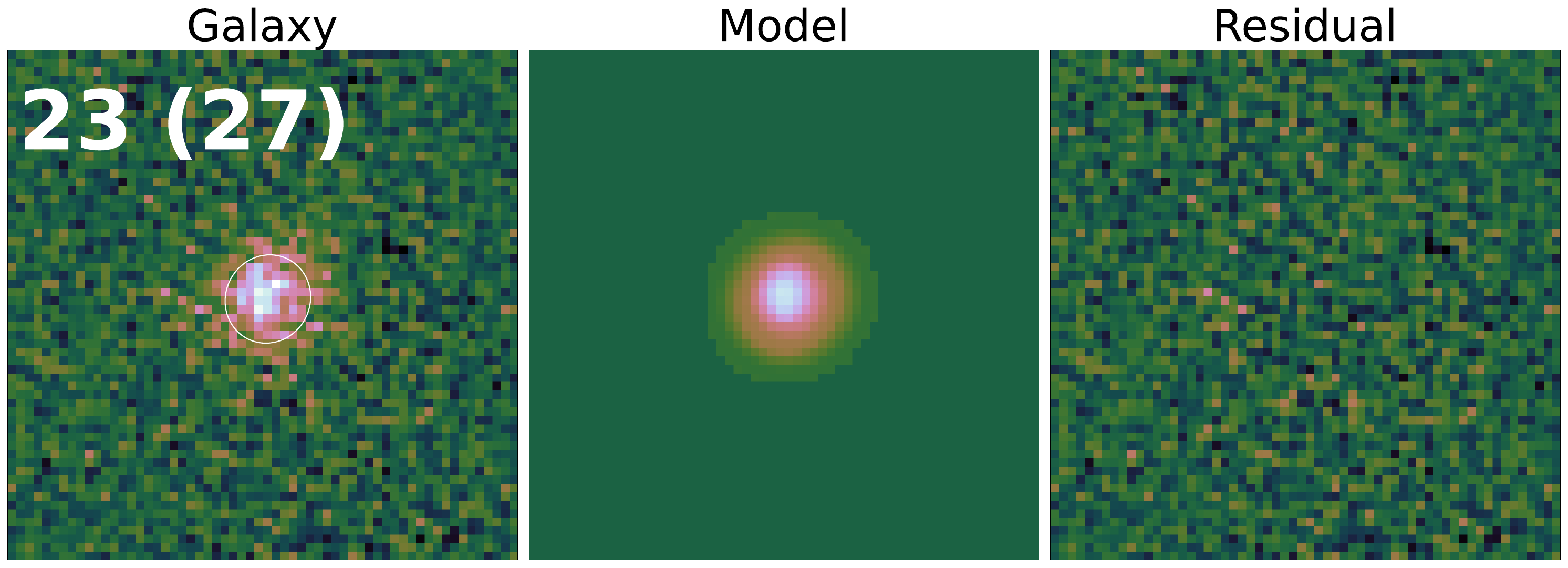}
    \end{minipage}\hfill

    \vspace{0.5em}

\end{figure*}

\bigskip


\begin{figure*}
    \centering
    \includegraphics[scale = 0.5]{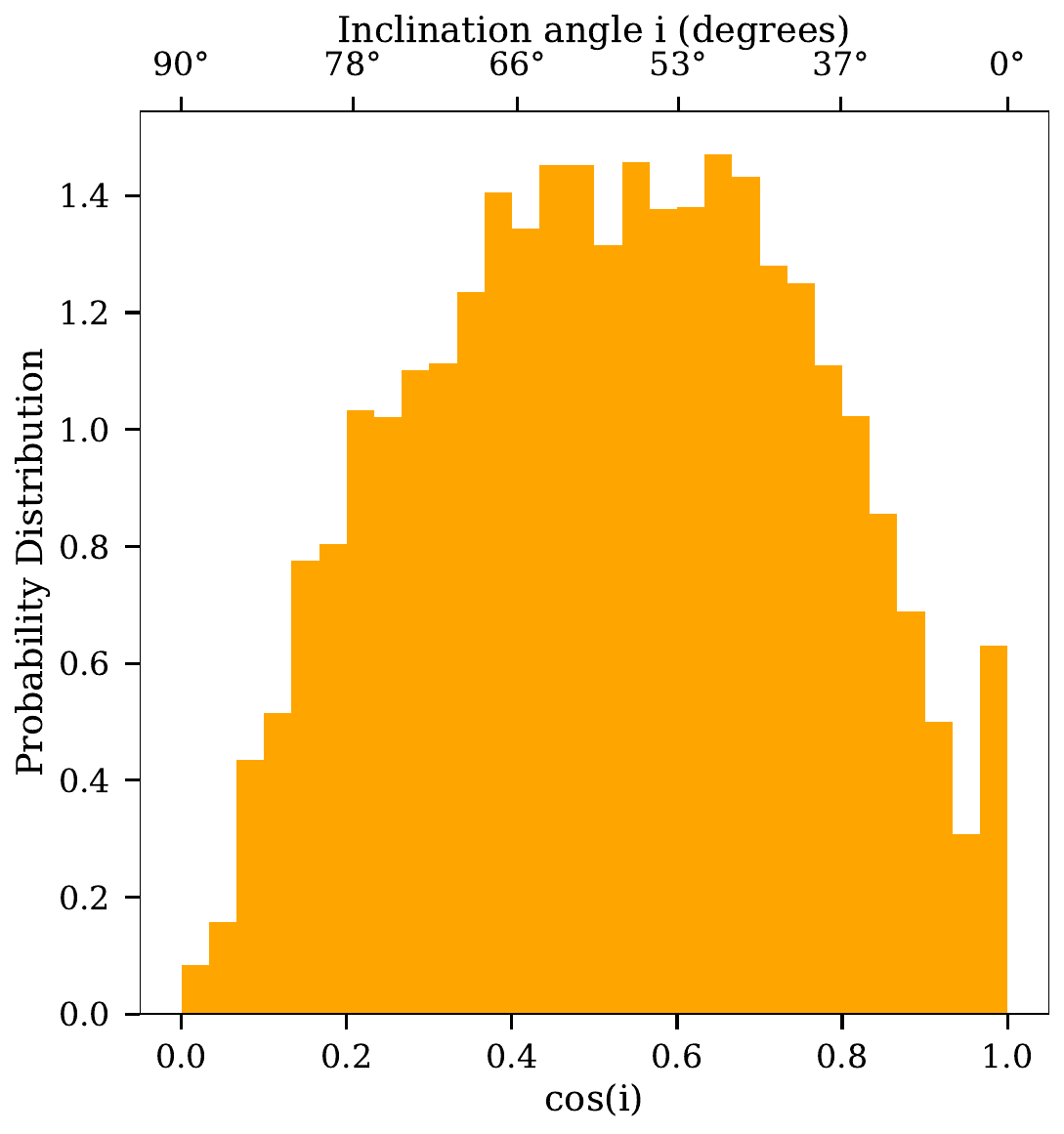}
    \caption{Histogram illustrating the cos($i$) of galaxies sampled from the SDSS-DR16 catalogue, adhering to the criteria outlined in the Methods section.}
    \label{fig:sdss_hist_axis_symmetric}
\end{figure*}




\bigskip

\begin{figure*}
    \centering
    \includegraphics[scale=0.45]{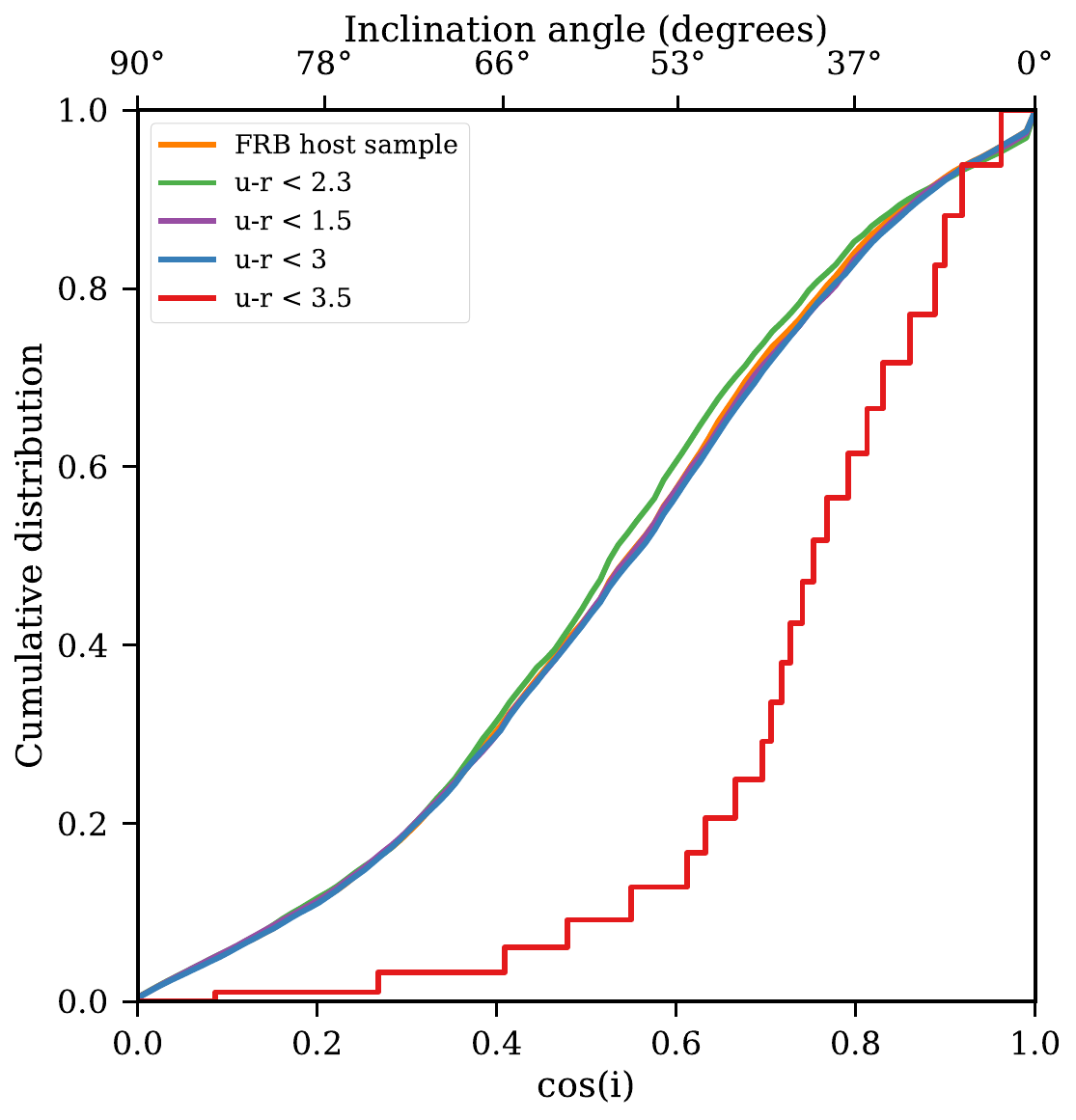}\\
    \includegraphics[scale=0.45]{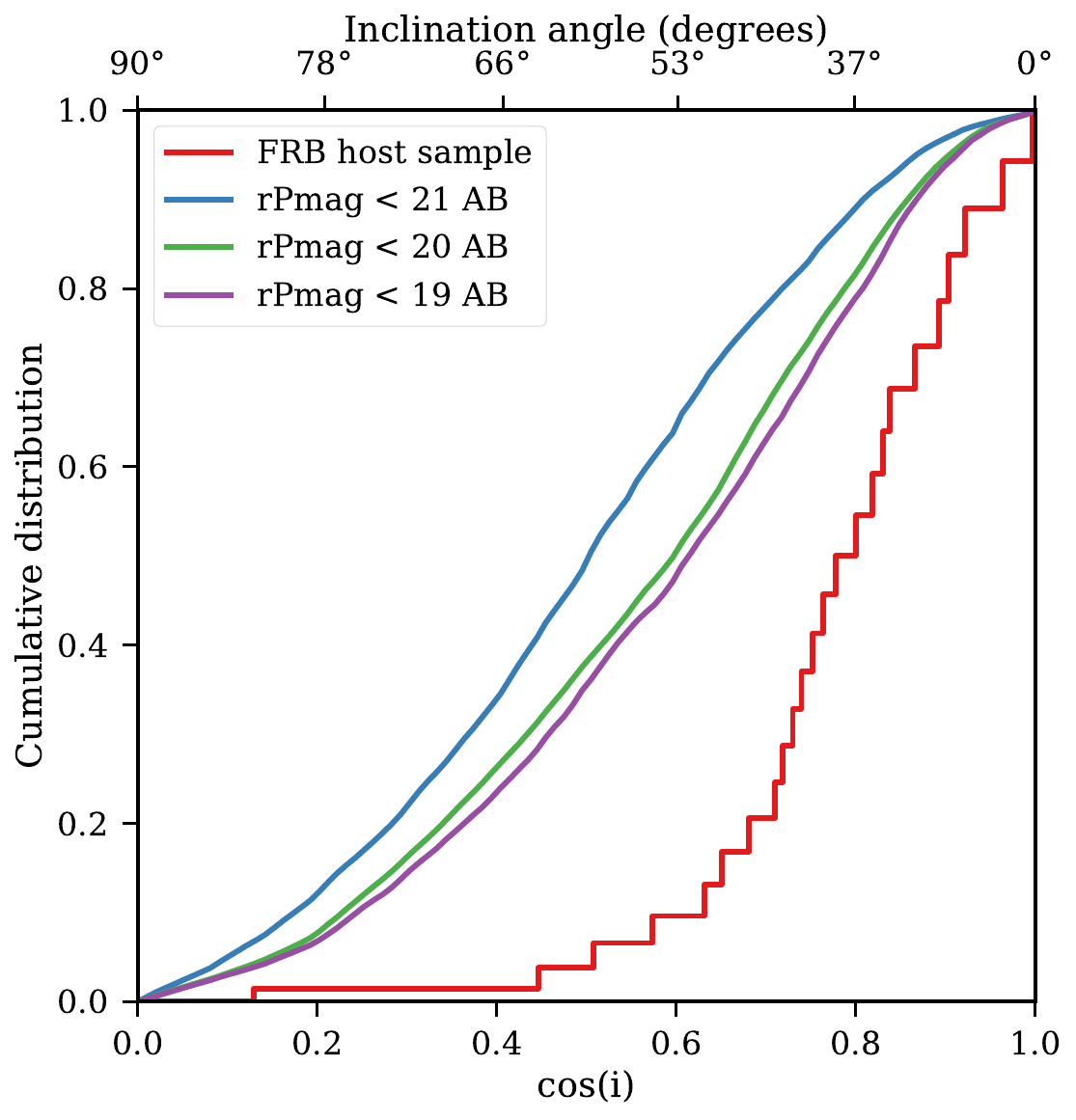}
    \includegraphics[scale = 0.45]{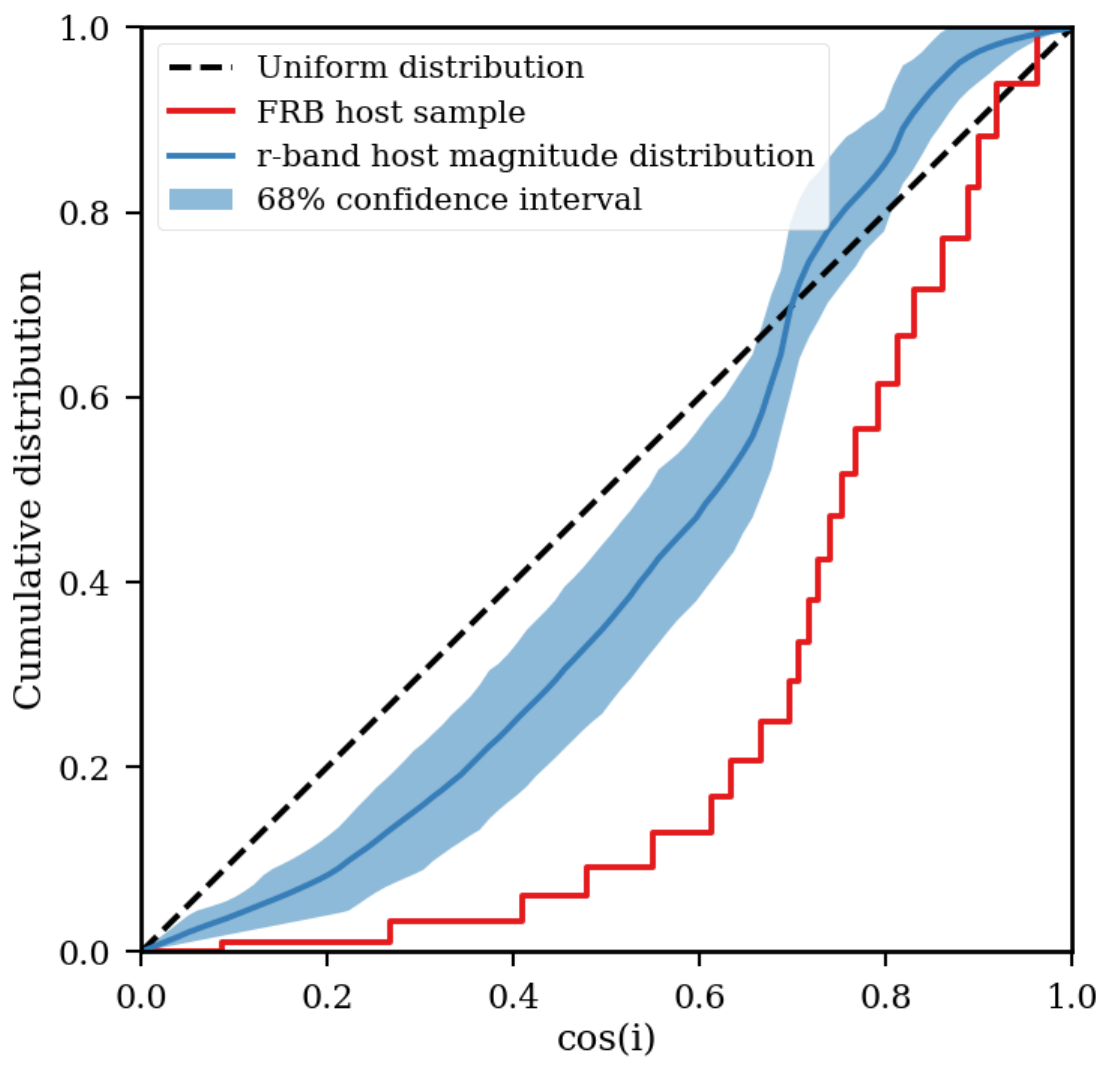}
    \caption{The top plot shows the mean $\cos(i)$ CDF for randomly sampled SDSS galaxies, employing a methodology similar to that used for Figure \ref{fig:main_cdf}, albeit with variations in u$-$r color thresholds. Similarly, the bottom left plot illustrates the impact on the mean $\cos(i)$ CDF of SDSS galaxies for different m$_{\rm r}$ thresholds. Finally, the bottom right plot displays the $\cos(i)$ CDF where SDSS galaxies are sampled in each iteration to match the m$_{\rm r}$ distribution of our FRB host sample. The blue shaded region represents the 68\% credible bound on the SDSS galaxy CDF to account for the small sample size. In all cases, the SDSS galaxy CDFs are found to be statistically different from the $\cos(i)$ CDF of FRB hosts in our sample, depicted in red in all three plots.}
    \label{fig:sdss_biases}
\end{figure*}




\end{document}